\documentclass[a4paper,11pt]{article}

\usepackage{jcappub} 
\usepackage{lineno}
\usepackage{longtable}
\usepackage{supertabular}
\usepackage{booktabs}
\usepackage{multirow}
\usepackage[utf8]{inputenc}   
\DeclareUnicodeCharacter{2212}{-}
\usepackage{hyperref}
\usepackage[T1]{fontenc}
\usepackage{appendix}
\usepackage{amsmath,amssymb}
\usepackage{graphicx}
\usepackage{dcolumn}
\usepackage{booktabs}
\usepackage{multirow}
\usepackage{bm}
\usepackage{hyperref}
\usepackage{braket}
\usepackage{soul}
\usepackage{pgfplotstable}
\usepackage{booktabs} 

\newcommand{\dd}{\mathrm{d}}

\usepackage{xcolor}     

\title{\boldmath Coupled by Design: Computing Kerr–Newman Quasinormal Modes with a Hybrid SpectralPINN Solver}

\author[]{Alexandre M. Pombo}

\affiliation[a]{CEICO, Institute of Physics of the Czech Academy of Sciences, Na Slovance 2, 182 21 Praha 8, Czechia}

\emailAdd{pombo@fzu.cz}

\abstract{We extend our \texttt{SpectralPINN} solver to the computation of Kerr-Newman quasinormal modes by applying it to solve the system of two coupled master PDEs -- advancing from the single, separable equation of the uncharged Kerr limit to a genuinely coupled two-field problem. The coupling between the gravitational and electromagnetic fields gives rise to two families of solutions: the photon-sphere, connecting with Kerr; and near-horizon, disconnected from the Kerr limit, each with two branches of solutions depending on the leading field: the gravitational- and vector-led. Benchmarking against publicly available datasets shows relative frequency errors of $\sim 10^{-4}$ worst case and $\sim10^{-7}$ for most cases. The computed public dataset spans five photon-sphere modes, both gravitational- and vector-led, up to $\ell=4$, as well as two fundamental gravitational-led near-horizon modes. The vector-led photon-sphere branch is computed and systematically characterized for the first time. We apply the dataset and observe the onset of the eigenvalue repulsion reported by Dias \textit{et al.}, to exclude an inter-polarization repulsion within the resolved domain, and to forecast Einstein Telescope constraints on the black-hole charge-to-mass ratio.}

\begin{document}

\maketitle

\flushbottom
%
\section{Introduction}\label{S1}
%
    Gravitational-wave (GW) observations by the LIGO--Virgo--KAGRA network \cite{LVK_open_data_O4a} (LVK) have transformed compact-object mergers into precision probes of the strong-field regime, with an ever-growing number of binary black hole (BH) and neutron star mergers documented in the GWTC catalogs~\cite{GWTC1,GWTC2,GWTC3,GWTC4,GWTC5}. 
    
    During the ringdown phase, the remnant radiates GWs whose time-domain signal is well-described by a superposition of quasinormal modes (QNMs) -- discrete complex frequencies that probe the remnant's linear stability and act as fingerprints of its nature. Recently, the LVK network detected GW250114, the loudest BH binary signal to date -- with a signal-to-noise ratio (SNR) $\sim 80$ --, whose post-merger data contains, besides the fundamental $(n,\ell, m) = (0,2,2)$ mode, an overtone mode at the $4.1\sigma$ level, consistent with the first overtone $(n,\ell, m) = (1,2,2)$ of a Kerr remnant~\cite{LVK:2025_GW250114_KerrArea,LVK:2026_GW250114_Spectroscopy} -- establishing overtone spectroscopy of a single harmonic at current sensitivity. However, full mode-resolved tests of gravity require ringdown SNRs of $\mathcal{O}(100)$ or higher~\cite{Bhagwat2020}. Next-generation observatories like the Einstein Telescope (ET)~\cite{Punturo2010ET}, Cosmic Explorer (CE)~\cite{Evans2021CE,Hall2022CE}, and LISA~\cite{LISA2017} are expected to achieve these SNRs for the first three individual modes: $(n,\ell,m) = \big[(0,2,2),\, (0,3,3),\, (0,4,4)\big]$. Such precision ringdown spectroscopy will allow us to quantify deviations from General Relativity, extra fields (\textit{e.g.,}~\cite{Isi2021}), or the presence of non-Kerr compact objects.

    Theoretically, while standard Kerr QNMs are cleanly obtained by separating the Teukolsky master equation~\cite{teukolsky1973perturbations,Berti2006}, computing QNMs for non-Kerr rotating compact objects remains challenging. These scenarios require robust predictions in settings where perturbations couple and standard separability fails. A particularly clean and instructive example is the electrically charged, spinning, Kerr--Newman (KN) BH solution~\cite{dias2015linear,mark2015quasinormal}. Even though astrophysical BHs are functionally uncharged, KN provides a crucial, minimal extension of Kerr where gravitational and electromagnetic perturbations couple, offering a controlled arena to understand mode mixing and spectral reorganizations. This coupling generates two distinct branches of solutions -- gravitational-led (Grav-led) and electromagnetic-led (Vect-led)\footnote{The direct coupling causes electromagnetic (gravitational) radiation to source gravitational (electromagnetic) radiation.} -- alongside two gravitationally relevant families: a photon-sphere/light-ring (PS) family connected to standard Kerr QNMs, and a near-horizon (NH) family prominent near extremality~\cite{yang2013quasinormal,yang2013branching,zimmerman2016damped,dias2022eigenvalue}. The NH family has no counterpart in the sub-extremal Kerr spectrum, driving novel phenomena like eigenvalue repulsion~\cite{dias2022eigenvalue} and resonances~\cite{yang2025black}, making KN a stringent benchmark for numerical solvers targeting coupled systems.

    In this manuscript we will extend \texttt{SpectralPINN} (first presented in~\cite{pombo2026teukolsky}) to compute the KN QNMs. \texttt{SpectralPINN} -- a hybrid pseudo-spectral/physics informed neural network (PINN) solver -- replaces the standard neural activation functions with Chebyshev polynomials of the first kind, enabling the network to act simultaneously as a spectral interpolator and as a PINN enforcing the PDE residuals and eigenvalue conditions. Boundary/normalization conditions are enforced through hard analytic masks encoding the horizon and infinity asymptotic behaviors (\textit{aka} hard enforcement). The obtained QNM frequencies reach a maximum combined $(\mathbb{R}+\mathbb{I})$ relative difference $\sim 10^{-4}$ (minimum $\sim 10^{-9}$) against the publicly available library \cite{berti2009ringdown} and published results~\cite{leaver1990quasinormal,kokkotas1988black,matyjasek2017quasinormal}. 
    
    To illustrate the observational impact of obtaining multiple modes accurately, we perform a one-dimensional scan in the charge-to-mass ration ($Q/M$), tracking the complex frequencies $\omega(Q)$ across the branch and extracting charge susceptibilities $d\omega/d(Q/M)^2$ -- the first-order response of the frequency to charge. These susceptibilities provide a compact bridge between theory and inference: they quantify how a measured ringdown frequency would shift under small departures from Kerr and therefore control the attainable constraints on $Q/M$. We translate the computed charge response into forecast constraints on $Q/M$ using the Einstein Telescope (ET-D) sensitivity curves with a two- and multi-mode analysis. Crucially, because different modes exhibit markedly different charge dependence, combining them breaks degeneracies and substantially improves constraints relative to two-mode analyses. 
    
    The computed dataset, containing both the Grav- and Vect-led PS $(n,\ell,m)=\big[$(0,2,2), (0,2,0), (1,2,2), (0,3,3), (0,4,4)$\big]$ modes and the Grav-led NH $(n,\ell,m)=\big[(0,2,2),\,(0,2,0)\big]$ is made publicly available, as well as the generating equations \cite{Zenodo}. 

    The paper is organized as follows. Sec.~\ref{S2} reviews the KN system of master equations. Sec.~\ref{S3} details the \texttt{SpectralPINN} solver, highlighting modifications from previous iterations, and benchmarks the updated framework against known results. Sec.~\ref{S4} applies the computed dataset to observe the onset of eigenvalue repulsion (Sec.~\ref{S42}) and forecast constraints using the ET--D sensitivity curve (Sec.~\ref{S43}). Finally, Sec.~\ref{S5} provides concluding remarks.
    
    Unless otherwise stated, throughout the text we consider geometrized units where $G=c= 1$, with $G$ the Newton's constant and $c$ denotes the speed of light. Both the charge, $Q$, and spin, $a$, parameters are in units of the mass, $M$. The signature of the spacetime is always $(-,+,+,+)$. We denote the real and imaginary parts as $\mathbb{R}$ and $\mathbb{I}$.
%
\section{Kerr-Newman QNM master equation}\label{S2}
%
    Following \cite{dias2022eigenvaluef}, the standard KN metric in Boyer--Lindquist coordinates $(t,r,\theta,\varphi)$ is given by,
    \begin{equation}
        ds^2 = -\frac{\Delta}{\Sigma}\big(\dd t-a\sin ^2 \theta \dd\varphi\big)^2+\Sigma\bigg(\frac{\dd r^2}{\Delta}+\dd \theta ^2\bigg)+\frac{\sin ^2 \theta }{\Sigma}\bigg[\big(r^2+a^2\big)\dd \varphi- a\dd t\bigg]^2  ,
    \end{equation}
    with the two metric functions $\Delta = r^2-2Mr+a^2+Q^2$ and $\Sigma = r^2+a^2\cos ^2 \theta$, and the Maxwell $4-$vector potential defined as
    \begin{equation}
        A_\mu = \frac{Q\, r}{\Sigma}\Big[\dd t-a\sin ^2 \theta\, \dd \varphi  \Big]\ .
    \end{equation}
    The BH solution is characterized by its mass, $M$, electric charge, $Q$, and the rotation parameter, $a$, which parametrizes the intrinsic angular momentum $J = M a$. The zeros of the $\Delta$ function determine the inner and outer horizon positions
    \begin{equation}
        r_\pm = M\pm \sqrt{M^2-a^2-Q^2}\ .
    \end{equation}
    The horizon's position radicand imposes the constraint on the maximum value of the $(a,Q)$ for a given $M$: $M^2\geqslant a^2+Q^2$. At the equality, $M^2 = a^2+Q^2$ and $r_+ = r_-$, where the regular KN BH reaches an extremal configuration. Physically, one is interested in the region outside the outer horizon ($r\geqslant r_+$) of sub-extremal KN BHs.
    
    The outer horizon is a Killing horizon generated by the Killing vector,
    \begin{equation}
        K = \partial_t+\Omega _H\, \partial_\varphi\ ,
    \end{equation}
    where the horizon angular velocity, $\Omega _H$, and temperature, $T_H$, are defined as,
    \begin{equation}
        \Omega_H = \frac{a}{r_+ ^2+a^2}\ , \qquad \qquad T_H = \frac{1}{4\pi r_+}\frac{r_+^2-a^2-Q^2}{r_+^2+a^2}\ .
    \end{equation}
    By virtue of being a Petrov type-D spacetime, the linear gravito-electromagnetic perturbations of a KN BH metric can be computed through the Newman--Penrose formalism just like the Teukolsky procedure for the Kerr solution~\cite{teukolsky1973perturbations,kokkotas1999quasi,ishak2018gravitational}. The QNM of the KN BHs are encoded in two coupled Teukolsky-like master equations first derived by Chandrasekhar \cite{chandrasekhar1985mathematical}, reformulated in a gauge compatible with numerical treatment by Mark~\textit{et al.}~\cite{mark2015quasinormal}, and then first numerically solved by Dias~\textit{et al.}~\cite{dias2015linear} (see also \cite{dias2022eigenvaluef} for a detailed derivation). The result is a system of two PDEs for the gravitational, $\psi_{-2}$, and electromagnetic, $\psi_{-1}$, gauge-invariant Newman--Penrose scalars that encode the Weyl and Maxwell fields~\cite{mark2015quasinormal,dias2015linear}:
    \begin{align}\label{E2.6}
        &\Big(\mathcal{F}_{-2}+Q^2 \mathcal{G}_{-2}\Big)\psi_{-2}+Q^2\mathcal{H}_{-2}\psi_{-1}=0\ ,\\
        &\Big(\mathcal{F}_{-1}+Q^2 \mathcal{G}_{-1}\Big)\psi_{-1}+Q^2\mathcal{H}_{-1}\psi_{-2}=0\ ,
        \label{E2.7}
    \end{align}
    where the second order differential operators $\{\mathcal{F},\mathcal{G},\mathcal{H}\}$ are given by,
    \begin{align}
        \mathcal{F}_{-2} & = \Delta \mathcal{D}^{\dagger}_{-1}\mathcal{D}_{0}+\mathcal{L}_{-1}\mathcal{L}^{\dagger}_{2}-6i\omega R\ ,\nonumber\\
        \mathcal{G}_{-2} & = \Delta \mathcal{D}_{-1} ^{\dagger}\alpha_- R^* \mathcal{D}_0 -3\Delta\mathcal{D}_{-1}^\dagger\alpha_--\mathcal{L}_{-1}\alpha_+R^*\mathcal{L}_2^\dagger+3\mathcal{L}_{-1}\alpha_+ia\sin \theta\ ,\nonumber\\
        \mathcal{H}_{-2} &= -\Delta \mathcal{D}_{-1}^\dagger\alpha_-R^*\mathcal{L}_{-1}-3\Delta\mathcal{D}_{-1}^\dagger\alpha_-ia\sin\theta -\mathcal{L}_{-1}\alpha_+R^*\Delta \mathcal{D}_{-1}^\dagger -3\mathcal{L}_{-1}\alpha_+\Delta \ ,\nonumber\\
        \mathcal{F}_{-1} & = \Delta \mathcal{D}_1\mathcal{D}_{-1}^\dagger +\mathcal{L}_2^\dagger \mathcal{L}_{-1}-6i\omega R\ ,\nonumber \\
        \mathcal{G}_{-1} & = -\mathcal{D}_{0}\alpha_+R^*\Delta \mathcal{D}_{-1}^\dagger -3\mathcal{D}_0\alpha_+\Delta +\mathcal{L}_2 ^\dagger\alpha_-R^*\mathcal{L}_{-1}+3\mathcal{L}_2^\dagger\alpha_-ia\sin \theta \ ,\nonumber\\
        \mathcal{H}_{-1} & = -\mathcal{D}_0 \alpha_+R^*\mathcal{L}_2 ^\dagger+3\mathcal{D}_0 \alpha_+ia\sin \theta-\mathcal{L}_2^\dagger\alpha_-R^*\mathcal{D}_0+4\mathcal{L}_2^\dagger\alpha_-\ .
    \end{align}
    Here $R= r+ia\cos \theta$ and $\alpha_\pm = \frac{1}{3\big(R^2 M-R Q^2\big)\pm Q^2 R^*}$, while the radial, $\mathcal{D}_i$, and angular, $\mathcal{L}_j$, Chandrasekhar operators are given by
    \begin{align}
        &\mathcal{D}_j = \partial_r+ \frac{iK_r}{\Delta}+2j\frac{r-M}{\Delta}\ ,\qquad \qquad K_r = am-\big(r^2+a^2\big)\omega\ ,\nonumber\\
        &\mathcal{L}_j = \partial_\theta +K_\theta +j\cot \theta\ , \qquad \qquad \qquad K_\theta = \frac{m}{\sin \theta}-a\omega \sin \theta\ .
    \end{align}
    The $\mathcal{D}_j^\dagger$ and $\mathcal{L}_j^\dagger$ are the complex conjugate Chandrasekhar operators which are obtained by $K_r \to -K_r$ in $\mathcal{D}_j$, and $K_\theta \to -K_\theta $ in $\mathcal{L}_j$. In the limit $Q\to 0$, the system of equations decouple and reduces to the standard Teukolsky equations for the gravitational ($s=-2$) \eqref{E2.6} and vector ($s=-1$)~\eqref{E2.7} perturbations of a Kerr BH. In the non-rotating limit of ($a\to 0$), the system correctly recovers the Reissner--Nordstr\"om BH ($Q\neq 0$) and Schwarzschild ($Q=0$) cases, providing a non-trivial check of our implementation.

    \bigskip

    To solve the system of coupled PDEs, one must impose appropriate physical boundary conditions \cite{dias2015linear}. At the horizon causality demands that there are only ingoing waves in Eddington--Finkelstein coordinates,
    \begin{equation}
    \label{E2.10}
        \psi_{s} \big|_{r_+}= (r-r_+)^{-s-i\frac{\omega-m\Omega_H}{4\pi T_H}}\big[a_s(\theta)+\mathcal{O}(r-r_+)\big]\ ,
    \end{equation}
    while at spatial infinity one must impose a pure outgoing wave behaviour,
    \begin{equation}
    \label{E2.11}
        \psi_{s} \big|_{r\to \infty}= e^{i\omega r}r^{-(2s+1)+i\omega\frac{r_+^2+a^2+Q^2}{r_+}}\Big[b_s(\theta)+\mathcal{O}\big(r^{-1}\big)\Big]\ .
    \end{equation}
    At the poles, regularity dictates that the perturbation functions must behave as
    \begin{equation}
        \psi_{s} \big|_{y\to \pm 1}= (1+y)^{\frac{|m-s|}{2}}(1-y)^{\frac{|m+s|}{2}}\Big[A_s(r)+\mathcal{O}(1\pm y)\Big]\ ,
    \end{equation}
    where $y = \cos \theta$ is the angular compactified coordinate. Following our previous work and the work done in \cite{luna2023solving}, the boundary conditions are implemented in the perturbation functions through an analytical mask (hard enforcement):
    \begin{equation}
        \psi_{s} (r,y) = e^{\omega (a y+i r)}(1+y)^{\frac{|m-s|}{2}}(1-y)^{\frac{|m+s|}{2}}(r-r_-)^{\sigma_-}(r-r_+)^{\sigma_+}\chi_s ( x,y)\ ,
    \end{equation}
    with
    \begin{align}
          \sigma_- =-1-s+2iM\omega+i\frac{\omega \big(a^2+r_+^2\big)-a m}{r_+-r_-} \ ,\qquad \sigma_+= -s-i \frac{\omega \big(a^2+r_+^2\big)-a m}{r_+-r_-}\ .
    \end{align}
    Observe that the choice of boundary condition enforcement reduces to the Frobenius analysis \eqref{E2.10} and \eqref{E2.11} in each respective limit. The additional terms were added to make the function smoother in the bulk and do not alter the asymptotic behaviors. The radial coordinate, $r$, was compactified as $x=r_+/r$, such that the resulting radial domain $r\in [r_+, +\infty[$ becomes $x\in [0,1]$.

    The resulting system of equations in the perturbation functions $\chi_s$, can be cast as
    \begin{equation}\label{E2.15}
        G \equiv G_{\rm grav}+ Q^2G_{\rm emf}=0\ ,\qquad \qquad E \equiv E_{\rm emf}+ Q^2 E_{\rm grav}= 0\ ,
    \end{equation}
    where the equation $G$ and $E$ denote the equation resulting from \eqref{E2.6} and \eqref{E2.7}, respectively, after the hard boundary condition/regularization enforcement and coordinate compactification. The individual $G$ and $E$ components are further factorized by the perturbation functions and their derivatives as,
    \begin{align}\label{E2.16}
        G_{\rm grav} = &\big(Gg_{0}+Gg_{1}\,\omega+Gg_{2}\,\omega^2\big)\chi_{-2}+\big(Gg_{y,0}+Gg_{y,1}\,\omega\big)\partial_y \chi_{-2}+\big(Gg_{x,0}+Gg_{x,1}\,\omega\big)\partial_x \chi_{-2}\nonumber\\
        &+Gg_{xx,0}\,\partial_{xx} \chi_{-2}+Gg_{yy,0}\,\partial_{yy} \chi_{-2}\ ,\\
        G_{\rm emf} =& \big(Ge_{0}+Ge_{1}\,\omega+Ge_{2}\,\omega^2\big)\chi_{-1}+\big(Ge_{y,0}+Ge_{y,1}\,\omega\big)\partial_y \chi_{-1}\nonumber\\
        &+\big(Ge_{x,0}+Ge_{x,1}\,\omega\big)\partial_x \chi_{-1}+Ge_{xy}\,\partial_{xy}\chi_{-1}\ ,\\
        E_{\rm grav} = & \big(Eg_0 + Eg_1\,\omega \big)\chi_{-2}+\big(Eg_{y,0}+Eg_{y,1}\,\omega\big)\partial_y\chi_{-2}+Eg_x\, \partial_{x}\chi_{-2}+Eg_{xy}\,\partial_{xy}\chi_{-2}\ ,  \\
        E_{\rm emf} = & \big(Ee_{0}+Ee_{1}\,\omega+Ee_{2}\,\omega^2\big)\chi_{-1}+\big(Ee_{y,0}+Ee_{y,1}\,\omega\big)\partial_y\chi_{-1}+\big(Ee_{x,0}+Ee_{x,1}\,\omega\big)\partial_x\chi_{-1}\nonumber\\
        &+Ee_{yy}\,\partial_{yy}\chi_{-1}+Ee_{xx}\,\partial_{xx}\chi_{-1}\ ,\label{E2.19}
    \end{align}
    where the $Gg$, $Ge$, $Eg$ and $Ee$ are the remaining non-trivial factorized terms present in the coupled PDE system. It is important to stress that, due to its pseudo-spectral foundation, our numerical method, \texttt{SpectralPINN}, is particularly sensitive to sharp features in the integrands. For this reason all factorized terms have been algebraically simplified to remove spurious denominators before being passed to the solver. In general, the equation terms are too lengthy to display in full. As a representative example, consider $Ge_{xy}$,
    \begin{align}
        Ge_{xy} =& -4 x^3 \Big[x^2 \big(a^2+Q^2\big)-2 x+1\Big] \big(a^2 x^2 y^2+1\big) \Big[2 Q^2 x \big(1+2 i a x y\big)+3 \big(a x y-i\big)^2\Big]\nonumber\\
        &\times \Big[3 i (a x y-i)^2-2 Q^2 x \big(a x y-2 i\big)\Big](1-y)^{\frac{1}{2} \big(3-|m-2| +|m-1|\big)} \ .
    \end{align}
    The exponent of the $(1-y)$ term was chosen such that it keeps all the $Ge$ and $Gg$ functions without a denominator for all $m\geqslant 0$ (the same procedure was performed for the $Ee$ and $Eg$ terms). All used factorized functions can be seen in both \texttt{Mathematica} and \texttt{PyTorch} format at \cite{Zenodo}. 

    Finally, the coupled master equations \eqref{E2.15} constitute a homogeneous linear eigenvalue problem. Once the eigenvalue $\omega$ is fixed, the perturbation functions $\chi_{-1}$ and $\chi_{-2}$ satisfy a linear PDE system and are therefore defined only up to an overall multiplicative constant, $\chi_s \rightarrow C\,\chi_s$. 
    
    This scaling freedom is physically irrelevant -- QNM amplitudes are fixed by the initial data rather than by the mode equation itself -- but it introduces a degeneracy in any optimization-based solver, since an entire one-parameter family of eigenfunctions corresponds to the same eigenvalue. For this reason one must supplement the QNM equations with a normalization condition that fixes the overall scale of the mode. Following Leaver's original construction for Kerr BHs \cite{leaver1985analytic}, normalization is imposed  at a specific point in the compactified domain, namely: $\chi_s (1,-1) = 1$ (corresponding to the horizon at the south pole, where the mode is generically non-vanishing). The normalized component is chosen according to the leading field. Such normalization uniquely selects a representative within each equivalence class of rescaled eigenfunctions.

    As in the case of the boundary/regularity conditions, and from the study performed in \cite{pombo2026teukolsky}, the implemented hard normalization\footnote{While the translation normalization, $\chi_s (x,y) = 1+\big[\chi_s(x,y)-\chi_s(1,-1)\big]$, used in \cite{pombo2026teukolsky} is also compatible with the current setup, the ratio normalization has a faster numerical convergence.} is,
    \begin{equation}
       \chi_s \to \frac{\chi_s}{\chi_{-1} (1,-1)} \quad {\rm (Vect-led)}\qquad \qquad \chi_s \to \frac{\chi_s}{\chi_{-2} (1,-1)} \quad {\rm (Grav-led)}\ .
    \end{equation}
    Note that prescribing a normalization condition at any non-degenerate point also prevents the optimization solver from collapsing to the trivial zero solution.
%
\section{SpectralPINN}\label{S3}
%
    To numerically solve the coupled two-dimensional PDEs~\eqref{E2.15} describing the perturbation function, we use an in-house-developed solver, \texttt{SpectralPINN}, which combines a pseudo-spectral method, inspired by \cite{Blazquez2024}, with a PINN-like training strategy, inspired by \cite{luna2023solving}. A detailed explanation of the solver can be seen in \cite{pombo2026teukolsky}. In this section, we review the solver and describe the modifications introduced for the KN system. 
    
\begin{figure}[h!]
	   \centering
	   \includegraphics[width=1.0\textwidth]{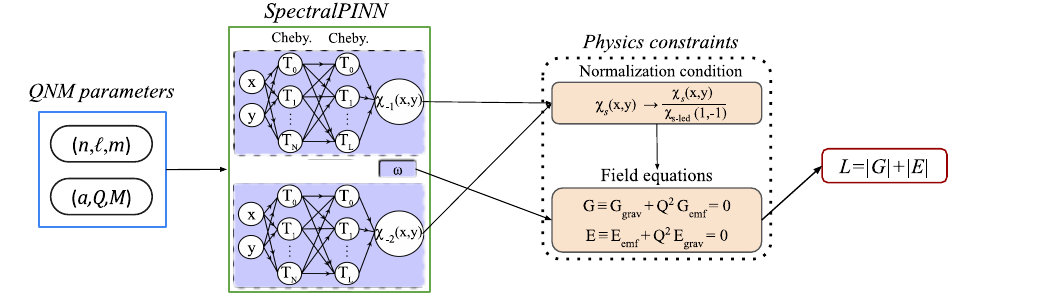}\\
	   \caption{Schematic representation of the \texttt{SpectralPINN} architecture. Each neuron corresponds to a Chebyshev basis function, and each ``layer'' is associated with a coordinate dependence. The compactified angular coordinate, $y$, is passed to the second block through an identity bypass, $K(y)=y$, that multiplies each radial basis evaluation, $T_i(x)$. The scheme is loosely inspired by the PINN designs in \cite{luna2023solving}.}
	   \label{F1}
    \end{figure}

    \texttt{SpectralPINN}'s basic key idea is to represent each perturbation function, $\chi_s$, in a basis of Chebyshev polynomials of the first kind, $T_n(x)$, 
    \begin{equation}\label{E3.2}
        \chi_s(x,y) = \sum_{i=0}^{N-1}\sum_{j=0}^{L-1} \mathcal{A}^s_{ij}\,T_i(x)\,T_j(y)\ ,
    \end{equation}
    and to treat the corresponding spectral amplitudes, $\mathcal{A}^s_{ij}$, as trainable parameters in a neural network optimization loop. The parameters $N,\, L$ correspond to the maximum orders retained in the radial and angular expansion, respectively. In practice, the ``layers'' of the network correspond to the coordinate directions, and the ``neurons'' are the basis functions $T_i(x)$ and $T_j(y)$. The only learnable parameters are the spectral amplitudes, $\mathcal{A}^s_{ij}$, and the eigenvalues $\omega$. A schematic representation of this architecture is shown in Fig.~\ref{F1}. This design sacrifices some of the flexibility of standard PINNs in exchange for precision and a strong inductive bias towards solutions compatible with the assumed basis symmetry.

    Due to the compactification used in Sec.~\ref{S2}, the angular dependence has already been mapped to $y\in[-1,1]$ and the explicit spin-weighted spherical-harmonic structure is encoded in the pre-factors. In this setting, expanding the angular dependence in Chebyshev polynomials leads to a more stable and accurate solver than using associated-Legendre polynomials directly, as measured by faster residual convergence at comparable resolution.

    Following standard spectral practice, the basis functions are evaluated on collocation grids matched to their orthogonality properties. The compactified radial coordinate $x\in[0,1]$, is mapped into a Chebyshev--Gauss--Lobatto grid, while the angular coordinate, $y\in [-1,1]$, is mapped into the usual Gauss--Lobatto nodes:
    \begin{align}\label{E3.3}
      x_k &=\frac{1}{2}\bigg[1-\cos\!\left(\frac{k\,\pi}{N_x-1}\right)\bigg]\ ,\qquad k=0,\, \dots\, ,N_x-1\ ,\nonumber\\
      y_q &=\cos\!\left(\frac{q \,\pi}{N_y-1}\right)\ ,\qquad \qquad \quad\ \,q=0,\, \dots\, ,N_y-1\ ,
    \end{align}
    with $N_x = \frac{3}{2}N$ and $N_y = \frac{3}{2}L$ the number of collocation points in the radial and angular directions, respectively. The $\frac{3}{2}$ factor was chosen to avoid aliasing. The grid points are fixed \textit{a priori} and depend only on $(N_x,\, N_y)$.

    The knowledge of the basis functions' analytic form and evaluation points allows us to pre-compute: \textit{i)} the Chebyshev values $T_i(x_k)$ and $T_j(y_q)$ and the corresponding differentiation matrices $D_x$, $D_{xx}$ (first and second derivatives in $x$), $D_y$, $D_{yy}$ (first and second derivatives in $y$) and $D_{xy}$ (cross derivative), obtained from the standard recurrence relations for Chebyshev polynomials~\cite{MasonHandscomb2003}; and \textit{ii)} all the coordinate-dependent coefficient functions appearing in the compactified functions \eqref{E2.16}--\eqref{E2.19}, \textit{i.e.,}\ $Gg,\ Ge,\, Eg,\ Ee$. Since the derivatives of the basis are known analytically, no additional numerical differentiation scheme is required, which simplifies the implementation and improves accuracy. These pre-computations also avoid redundant calculations at each training iteration and significantly reduce the computational cost.
    In each training cycle, the current set of amplitudes and eigenvalues is used to reconstruct the perturbation functions, $\chi_s$, and their derivatives on the collocation grid. After multiplying by the respective $\{Gg,\, Ge,\, Eg,\ Ee\}$ factors, the reconstructions are inserted into the operators $G$ and $E$ to obtain the residuals of the coupled PDE system. The residuals then enter the loss functional through a mean absolute error (MAE, L$_1$),
    \begin{equation}
        \mathcal{L}oss = \frac{1}{N_x N_y}\sum_{i,j} ^{N_x,N_y}\Big[|E|+|G|\Big]\ .
    \end{equation}
    
    In practice, moderate spectral resolutions provide an optimal balance between accuracy and trainability for the KN QNMs considered here; in all following computations, we take $N\times L = 30\times 30$ with $N_x \times N_y = 45\times 45$.
%
    \subsection{Training}\label{S31}
%
    In practice, the code runs transparently on either CPU or GPU; our benchmarks were obtained on NVIDIA A100 and P100 GPUs or a Ryzen 7 7840HS–class CPU (16 threads). On all hardware configurations, typical runs require $\mathcal{O}(1$–$3)$ minutes to reach a residual loss $\mathcal{L}oss \simeq 10^{-7}$ where it tends to stabilize\footnote{Because the governing equations enter directly in the loss, PINNs and \texttt{SpectralPINN} strongly mitigate issues common in standard neural networks, such as overfitting, poor generalization, and reliance on large labeled datasets, as the governing physics directly constrains the solution space.}.
    
    In this work we use a two-step optimization. In the first step, \texttt{SpectralPINN} uses the HybridAdamDecoupled (HAdamD) optimizer developed in \cite{pombo2026teukolsky} to locate the correct solution branch from a cold start and drive it to $\mathcal{L}oss\sim 10^{-5}$; in the second step, the solver uses a Newton--Raphson-like (NR) optimizer to drive it to $\mathcal{L}oss\sim 10^{-12}$. This dual-step training allows us to directly converge to any solution without requiring an evolution from a previous seed. The HAdamD stage drives the solution into the basin of the correct minimum, from which the NR refinement converges to machine-level accuracy floor for this problem.

    Since we have already described in a previous work~\cite{pombo2026teukolsky} the HAdamD optimizer and associated scheduler in detail, we briefly review the overall process before focusing on the new NR-like optimizer. 
 
        \subsubsection{Hybrid Adam Optimizer}\label{S311}

    A key ingredient in our training procedure is a stable optimizer that can handle complex-valued parameters. Many of the trainable quantities in our models (\textit{e.g.,} the spectral amplitudes and eigenvalues) are complex, and naively applying the standard real-valued \texttt{Adam} optimizer to their real and imaginary parts can lead to sub-optimal convergence. We therefore implement a custom first-order optimizer, \texttt{HAdamD}, which generalizes \texttt{Adam} to the complex domain while keeping a simple interface.
    
    \texttt{HAdamD} treats complex parameters as 2D real vectors but preserves a genuinely complex representation for the first moment. Let $u_t \in \mathbb{C}^{n}$ be a vector of complex numbers with gradients $U_t \in \mathbb{C}^n$ defined as,
    \begin{equation}
        u_t = u_t ^{\mathbb{R}}+ i\, u_t ^{\mathbb{I}}\ , \qquad U_t = U_t ^{\mathbb{R}}+ i\, U_t ^{\mathbb{I}} \ .
    \end{equation}
    The optimizer maintains a complex first moment $m_t \in \mathbb{C}^n$, and two decoupled second moments: $v_t ^{\mathbb{R}}$, $v_t ^{\mathbb{I}}\in \mathbb{R}^n$; tracking the variance along the real and imaginary axes separately. The moment updates are
    \begin{equation}
        m_t = \beta_1\, m_{t-1}+(1-\beta_1)\, U_t\ , \,\,\,\,\,\, v_t ^{\mathbb{R}} = \beta_2\, v_{t-1} ^{\mathbb{R}}+(1-\beta_2)\big(U_t^{\mathbb{R}}\big)^2\ ,\,\,\,\,\, v_t ^{\mathbb{I}} = \beta_2\, v_{t-1} ^{\mathbb{I}}+(1-\beta_2)\big(U_t^{\mathbb{I}}\big)^2\ .
    \end{equation}
    After standard \texttt{Adam} bias correction,
    \begin{equation}
        \hat{m}_t = \frac{m_t}{1-\beta_1 ^t}\ , \qquad \hat{v}_t ^{\mathbb{R}} =\frac{v_t^{\mathbb{R}}}{1-\beta_2 ^t}\ , \qquad \hat{v}_t ^{\mathbb{I}} =\frac{v_t^{\mathbb{I}}}{1-\beta_2 ^t}\ ,
    \end{equation}
    the update is applied axis-wise, with separated root-mean-square (RMS) of the gradients normalization for the real and imaginary parts:
    \begin{equation}
        \Delta u_t ^{\mathbb{R}} = \frac{\hat{m}_t ^{\mathbb{R}}}{\sqrt{\hat{v }_t ^{\mathbb{R} } }+\epsilon }\, ,\qquad \Delta u_t ^{\mathbb{I}} = \frac{\hat{m}_t ^\mathbb{I}}{\sqrt{\hat{v}_t ^{\mathbb{I}}}+\epsilon}\ ,\qquad u_{t+1} = u_{t}-\alpha \big[\Delta u_t ^{\mathbb{R}}+i \Delta u_t ^{\mathbb{I}}\big]\ .
    \end{equation}
    The direction of descent is still governed by a single complex first moment $m_t$, but the adaptive scaling is decoupled per axis, providing distinct learning rates along the real and imaginary directions. In practice, this improves stability for complex weights whose real and imaginary parts experience gradients of different magnitude or curvature.

    For purely real parameters, \texttt{HAdamD} reduces exactly to the standard \texttt{Adam}. In that case, \texttt{HAdamD} stores only a real first moment and reuses a single second-moment buffer, $v_t ^{\mathbb{R}}$, discarding the imaginary branch. As a result, a single optimizer instance can seamlessly handle models with arbitrary mixtures of real and complex tensors, using standard \texttt{Adam} updates on real-valued layers and our decoupled complex variant where needed, without changes to the training loop.

       \subsubsection{Newton-Raphson Refinement}\label{S312}

    The first-order \texttt{HAdamD} training described above typically drives the loss to $\mathcal{L}oss\sim 10^{-4}$, at which point the optimizer/scheduler combination reaches its capacity to further decrease the loss. To push the eigenvalue accuracy beyond the first-order plateau, we append a second-order ``Newton--Raphson'' (NR) refinement stage that is sensitive to the full curvature information over the parameters space. This second-order optimizer exploits the special structure of our problem: the loss depends on the eigenvalue $\omega$ through a known, differentiable residual whose curvature can be computed exactly via automatic differentiation. 

    The NR refinement starts by collecting all trainable parameters of the \texttt{SpectralPINN} $(\omega,\, \mathcal{A}^{-1}_{ij},\, \mathcal{A}^{-2}_{ij})\in \mathbb{C}$ into a single flat weight vector $\Theta \in \mathbb{R}^{n_\Theta}$, obtained by interleaving the real and imaginary parts of each complex parameter\footnote{For a spectral resolution $N\times L=30\times 30$, this gives $n_\Theta = 3602$ real trainable weights.}.

    Simultaneously, rather than reducing the coupled PDE residual to a single scalar loss via the MAE, we retain the per-collocation-point residuals as a vector-valued output. Letting $G_{kq}$ and $E_{kq}$ denote the complex gravitational and electromagnetic residuals \eqref{E2.15} at collocation point $(x_k, y_q)$, the network's output vector $\textbf{R}\in \mathbb{R}^{n_{R}}$ is assembled by stacking the real and imaginary parts of both physical sectors,
    \begin{equation}
        \textbf{R}(\Theta) = \Big(G_{00}^\mathbb{R},\, G_{00}^\mathbb{I},...,G_{N_x N_y}^\mathbb{R},\,G_{N_x N_y}^\mathbb{I},\, E_{00}^\mathbb{R},\, E_{00}^\mathbb{I},...,\, E_{N_x N_y}^\mathbb{R},\, E_{N_x N_y}^\mathbb{I}\Big)^{T}\ ,
    \end{equation}
    giving $n_R = 2\times2\times N_x\times N_y $ real components (for $N=L=30$ and $N_x=N_y=45$ we have $n_R = 8100$). A perfectly trained network corresponds to $\textbf{R}(\Theta)=0$. After the first-order training the network is already near the basin of attraction of a second-order method, strongly favouring convergence of the latter.

    In the small residual regime reached after the \texttt{HAdamD} pre-training, the Gauss--Newton approximation to the Hessian becomes increasingly accurate, and the optimal weight update reduces to
    \begin{equation}
        \delta \Theta_t = -J_t ^+ \textbf{R}(\Theta _t)\ ,
    \end{equation}
    where $J_t = \frac{\partial \textbf{R}}{\partial \Theta_t}\in \mathbb{R}^{n_R\times n_\Theta}$ is the Jacobian of the network output with respect to all weights and $J_t ^+$ is the Moore--Penrose pseudo-inverse. Unlike the block coordinate \texttt{HAdamD} schedule, which must alternate between amplitude and eigenvalue blocks, the Gauss--Newton step updates all weights jointly: the full Jacobian encodes how a perturbation of any weight propagates through the spectral reconstruction and coupled PDE operators into every collocation-point residual, enabling large, coordinated steps that are inaccessible to first-order methods.

    The Jacobian is assembled via \texttt{PyTorch}'s functional automatic differentiation in vectorized mode -- tractable here because \texttt{SpectralPINN} has only $\mathcal{O}(10^3)$ trainable weights and a purely algebraic forward pass. To maintain a live autograd graph, the model's internal weights are temporarily replaced by differentiable views into $\Theta$ (a graph-aware weight injection analogous to the functional-parameter paradigm used in meta-learning and neural tangent kernel computations \cite{finn2017model,jacot2018neural}) and restored after each Jacobian call. The pseudo-inverse is computed via truncated singular value decomposition with $r_{\rm cond}=10^{-12}$, which acts as an implicit Tikhonov regularizer suppressing near-degenerate directions in weight space.

    The optimizer runs with unit learning rate, $\Theta_{t+1}=\Theta_t+\delta\Theta_{t}$, until convergence is reached, typically $5-10$ steps. Convergence is monitored through the residual norm $|R|\leqslant 10^{-12}$. At convergence, the weight vector is unpacked and committed back to the model, updating $\omega$, $\mathcal{A}^{-2}_{ij}$, and $\mathcal{A}^{-1}_{ij}$ simultaneously.

    To summarize, the full training pipeline for each point in the $(a,\, Q)$ parameter space proceeds in two stages. First, the block-coordinate first-order optimizer (\texttt{HAdamD}) with plateau-based scheduling trains the network until the amplitude learning rate reaches its floor of $10^{-14}$, typically requiring $\mathcal{O}(10^3\text{--}10^4)$ outer epochs and driving the residual loss to $\mathcal{L}oss \sim 10^{-4}$. Then, the second-order NR optimizer fine-tunes all weights jointly, driving the residual norm down to $\sim 10^{-12}$ in $5-10$ optimizer steps. The entire pipeline for a single $(a,\, Q)$ point runs in minutes on a single GPU or comparable CPU, making systematic parameter-space scans in $Q/M$ at fixed $a/M$ entirely feasible. 
    
\begin{figure}[h!]
	   \centering
	   \includegraphics[width=1\textwidth]{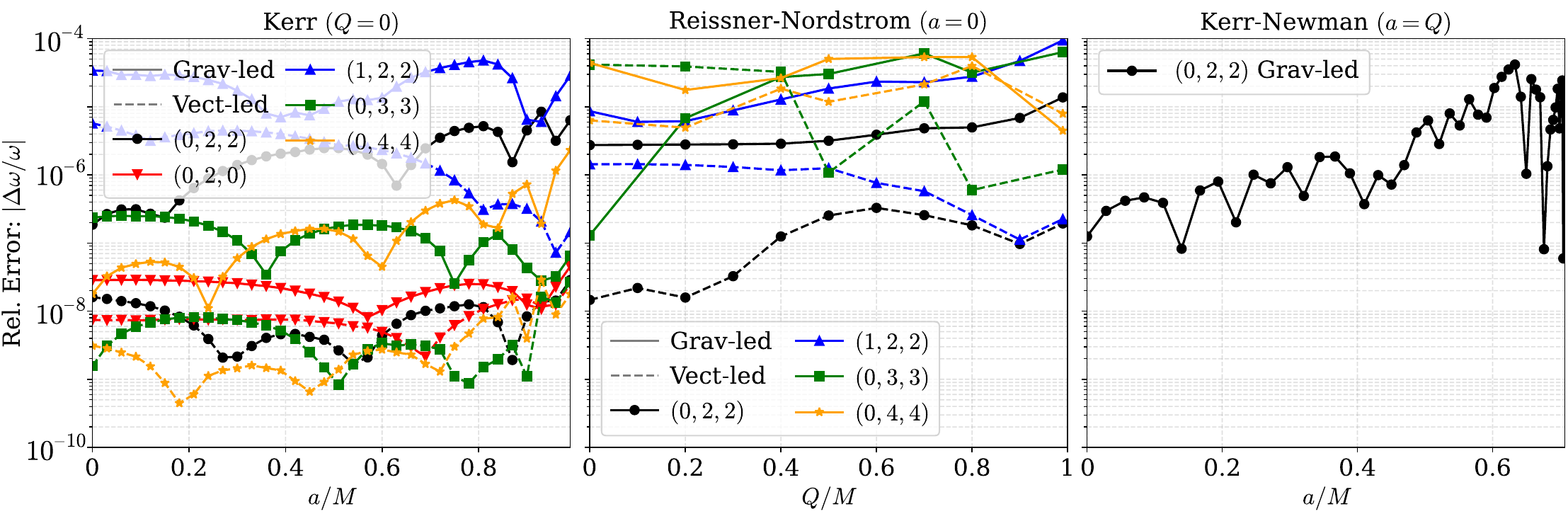}\\
	   \caption{Combined ($\mathbb{R}+\mathbb{I}$) relative error of the perturbation frequency $\omega$ computed with $N_x \times N_y = 45\times 45$ and $N\times L = 30\times 30$ for: (left panel) Kerr BH ($Q=0$ regime) with respect to the public data available at \cite{berti2009quasinormal,berti2009ringdown} as a function of $a/M$; (middle panel) Reissner--Nordstr\"om BH ($a=0$ regime) with respect to values computed in \cite{leaver1990quasinormal,kokkotas1988black,matyjasek2017quasinormal} as a function of $Q/M$; (right panel) Kerr--Newman diagonal ($Q=a$ regime) with respect to the public values computed by \cite{dias2015linear,berti2009ringdown} as a function of $Q/M=a/M$. Results are shown for both the Grav-led (solid lines) and Vect-led (dashed line) modes.}
	   \label{F2}
    \end{figure}

    To demonstrate the accuracy of our results, we plot in Fig.~\ref{F2}, the combined ($\mathbb{R}+\mathbb{I}$) relative error of the frequency obtained through the \texttt{SpectralPINN} against the publicly available results, namely, the Kerr $(Q=0)$ regime \cite{berti2009quasinormal,berti2009ringdown} (left panel); the Reissner--Nordstr\"om $(a=0)$ regime \cite{leaver1990quasinormal,kokkotas1988black,matyjasek2017quasinormal} (middle panel); and the Kerr--Newman $a=Q$ diagonal line \cite{dias2015linear,berti2009ringdown} (right panel). The three panels of Fig.~\ref{F2} confirm that the HAdamD+NR pipeline reproduces known QNM frequencies with a relative difference that spans from $\sim 10^{-9}$ to $\sim 10^{-4}$, with the largest errors confined to near-extremal regimes and overtones. These figures represent an improvement of roughly three orders of magnitude over the HAdamD-only results reported in~\cite{pombo2026teukolsky} for the Kerr case.
%
\section{Kerr-Newman Quasinormal Modes}\label{S4}
%
    With \texttt{SpectralPINN} validated (see Fig.~\ref{F2}), we now present and analyze the full dataset of KN QNM frequencies computed in this work. As introduced in Sec.~\ref{S1}, each set of quantum numbers $(n,\ell,m)$ admits both a Grav-led and Vect-led solution branch, and modes are further classified into the PS (photon-sphere) and NH (near-horizon) families. The results are organized into three subsections: the scope and structure of the dataset, the onset of eigenvalue repulsion between the PS and NH families, and the Einstein Telescope forecast for charge constraints derived from the computed spectra.    
%
    \subsection{Dataset}\label{S41}
%
    We compute the gravito-electromagnetic QNM spectrum for the KN BH across the two-dimensional sub-extremal parameter space $(a/M, Q/M)$. For the PS Grav- and Vect-led modes, the computation is organized as a series of one-dimensional charge scans at fixed spin: for each value of $a/M$, we sweep $Q/M$ from 0 up to approximately $99\%$ of the extremality bound $Q_{\rm max}/M = \sqrt{1-a^2/M^2}$, beyond which the solver ceases to converge reliably for the chosen number of basis polynomials, collocation numbers and double precision (\texttt{float64}). The spin values are sampled uniformly at $a/M =0, 0.05,0.10,..., 0.90$ and then progressively finer spacing near extremality: $a/M=0.92,0.94,0.95,0.96,0.97,0.98,0.99,0.995,0.999$, giving $27$ spin-slices in total. Along each slice, the charge is sampled with $\mathcal{O}(10^5)$ points, yielding a dataset of $\mathcal{O}(10^6-10^7)$ frequency evaluations per mode. For the NH modes, due to their rapid increase in the $-\omega_\mathbb{I}$ component as one goes away from the extremal RN BH solution, the one-dimensional scan is performed in the spin while keeping the charge constant. In this case, starting near extremality, for each value of $Q/M \in [$0.9995,0.999,0.998,0.996,0.994, 0.992,0.990,0.980,0.970,0.960,0.950$]$, the spin is sampled with $\mathcal{O}(10^5)$ points, where the finer spacing is required to ensure reliable continuation of the QNM branch. 

    At each point $(a/M, Q/M)$, \texttt{SpectralPINN} returns the complex frequency $\omega = \omega_\mathbb{R}+i\omega _\mathbb{I}$, together with the two-dimensional eigenfunctions for both the gravitational and electromagnetic field perturbations. The choice of quantum numbers $(n,\ell,m)$ reflects the expected detectability hierarchy in next-generation ringdown observations: the $(0,2,2)$ mode is the dominant emission channel from quasi-circular binary mergers; the $(0,3,3)$ mode is the loudest and longest-lived sub-dominant mode, being followed by the $(0,4,4)$ which provides further leverage for high-mass-ratio or precessing systems; the $(0,2,0)$ mode becomes co-dominant for highly eccentric mergers; and the $(1,2,2)$ overtone is relevant when the ringdown analysis begins at or near the signal peaks. For the NH family, we focus on the $\ell = m = 2$ sector and the axisymmetric $m=0$ mode, for $n=0$, which are the NH modes most likely to interact with the dominant PS modes through eigenvalue repulsion.

\begin{figure}[h!]
	   \centering
	   \includegraphics[width=1.0\textwidth]{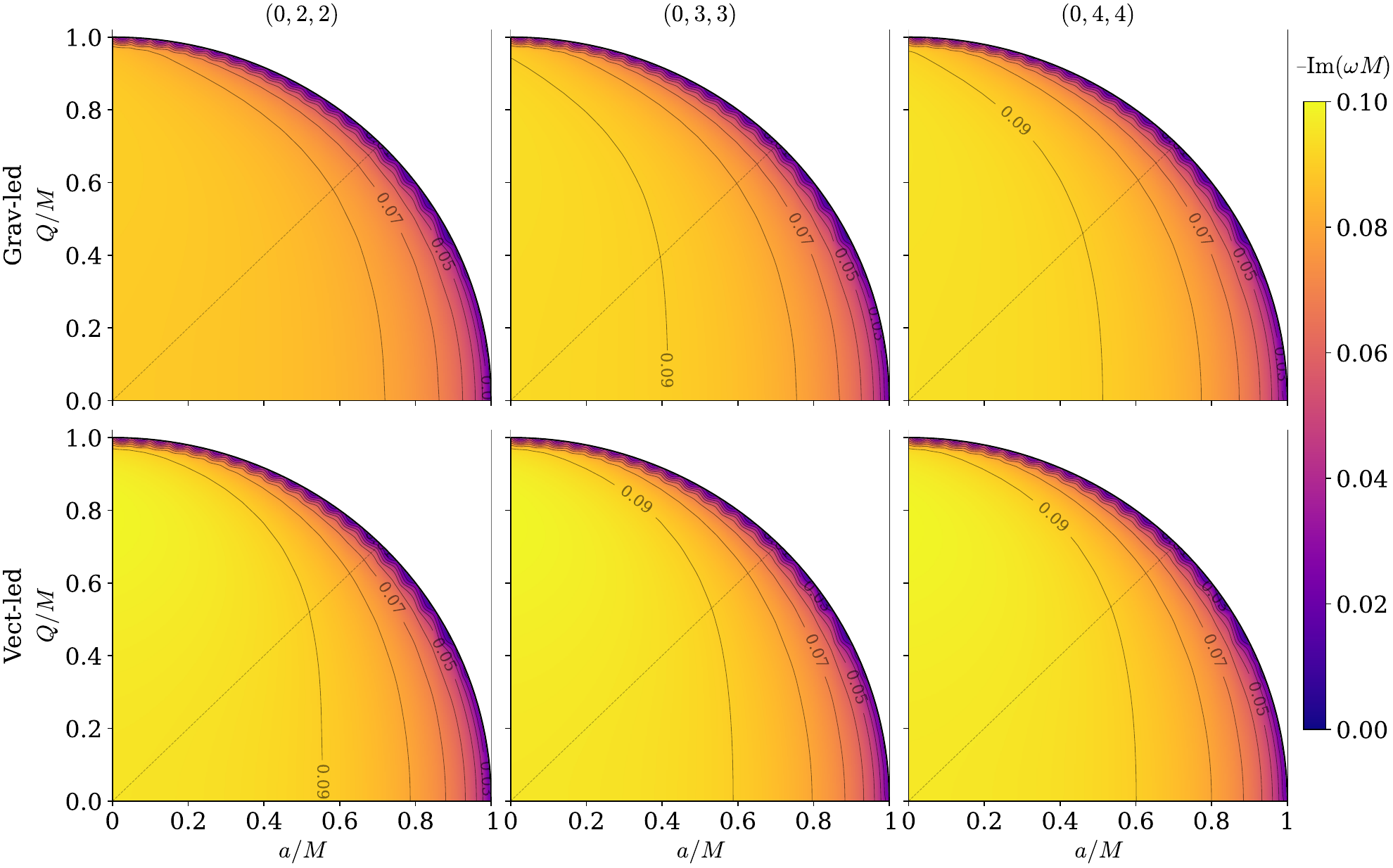}\\
	   \caption{Damping rate $-{\rm Im}(\omega\, M)$ interpolated heatmaps over the sub-extremal $(a/M, Q/M)$ parameter space for the three leading PS modes (left) $(0,2,2)$, (middle) $(0,3,3)$ and (right) $(0,4,4)$, for (top) Grav-led branch and (bottom) Vect-led branch.}
	   \label{F3}
    \end{figure}

    An example of the computed dataset can be seen in Fig.~\ref{F3} where we display the damping rate $-{\rm Im}(\omega\, M)$ as interpolated heatmaps over the sub-extremal $(a/M, Q/M)$ parameter space for the three leading PS modes $(0,2,2)$, $(0,3,3)$ and $(0,4,4)$. The top row shows the Grav-led branch and the bottom row the Vect-led branch. The domain is bounded from above by the extremality curve $Q/M=\sqrt{1-a^2 /M^2}$, along which the PS-family damping rates vanish as the modes approach the superradiant bound ${\rm Re}(\omega\, M)\to m\, \Omega_H$.
    The different response of the two polarizations to charge is a direct consequence of the coupled nature of the KN perturbation equations: the $Q^2$ mixing terms redistribute the perturbation energy between the gravitational and electromagnetic fields in a mode-dependent fashion. For instance, the $-{\rm Im}(\omega\, M)=0.07$ contour for the $(0,2,2)$ Grav-led mode sits at notably higher $Q/M$ than its Vect-led counterpart at the same spin. 
    
    However, we stress that, while the Vect-led branch is spectroscopically distinct from its Grav-led counterpart and constitutes an independent ringdown channel, its observational relevance is limited: in the $Q\to 0$ limit the Vect-led branch reduces to the purely electromagnetic Teukolsky mode, which does not couple to the gravitational waveform, and for $Q\neq 0$ its gravitational excitation is sourced only through the $Q^2$ mixing terms in~\eqref{E2.15}. Its contribution to the GW signal is, therefore, expected to be $Q^2$-suppressed relative to the Grav-led modes. We nonetheless include the Vect-led frequencies and damping rates in our public dataset, as they are required inputs for any such analysis and for studies of the coupled spectrum near extremality. The charge dependence of the Grav-led modes forms the basis of the multi-mode Fisher analysis in Sec.~\ref{S43}.

    Tracking the $n=1$ NH overtone to our target accuracy proved unfeasible in double precision near extremality — the continuation either locks onto the adjacent $n=2$ branch or the $−i/4$ near-extremal mode. A similar limitation occurs for the NH Vect-led modes. Following Dias \textit{et al.} \cite{dias2022eigenvalue}, who employ quadruple precision throughout, we defer this to future work.
%
    \subsection{Eigenvalue repulsion}\label{S42}
%
    A distinctive feature of the KN QNM spectrum, with no analogue in the Kerr or Reissner--Nordstr\"om individual limits, is the occurrence of eigenvalue repulsion between mode families. For Kerr BHs, the perturbation is described by a single master field equation, the Teukolsky equation; different mode families are eigenvalues of the same single-field system and can cross freely in the complex-frequency plane. For Reissner--Nordstr\"om, the gravitational and electromagnetic perturbations are coupled by the background charge, but the spherical symmetry of the spacetime allows the coupled system of ODEs to be reduced -- via Chandrasekhar transformation -- into independent master equations, so that the two sectors effectively decouple and crossings remain simple. In the KN system, neither simplification is available: spin breaks the spherical symmetry that permits separation into same-field modes, while charge introduces the coupling that spin alone does not produce. The resulting irreducible coupling of gravitational and electromagnetic perturbations through the $Q^2$ terms in \eqref{E2.15} generically lifts would-be degeneracies. When two modes approach each other in the complex plane, they repel rather than cross, exchanging their character in the process.

    This phenomenon was first identified in the KN context by Dias \textit{et al.} \cite{dias2022eigenvalue}, who studied the PS--NH interaction between the Grav-led $(0,2,2)$ and $(1,2,2)$ modes along the one-dimensional charge scan for fixed $a/a_{\rm ext}$. In the Reissner--Nordstr\"om limit ($a=0$), the PS and NH families are unambiguously distinct: the fundamental PS $(0,2,2)$ mode dominates at low charge while the fundamental NH $(0,2,2)$ mode takes over near extremality, with a simple crossover at $Q/r_+ \approx 0.959$. As spin increases, however, the crossing is replaced by an avoided crossing -- the two branches develop a frequency gap, exchange character, and reconnect into hybrid curves that can no longer be cleanly assigned to a single family. At still higher spins $(a\gtrsim 0.96\ a_{\rm ext})$, the PS and NH $(0,2,2)$ branches merge entirely into a single curve well described by the matched asymptotic expansion near extremality.

\begin{figure}[h!]
        \centering
        \includegraphics[width=0.49\textwidth]{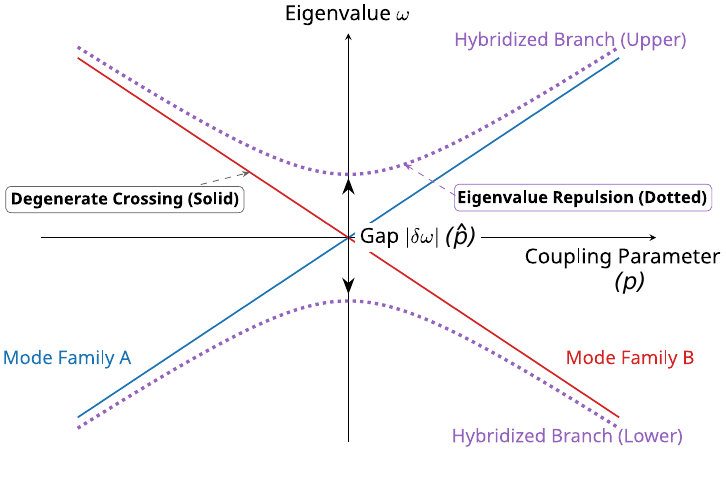}
        \includegraphics[width=0.5\textwidth]{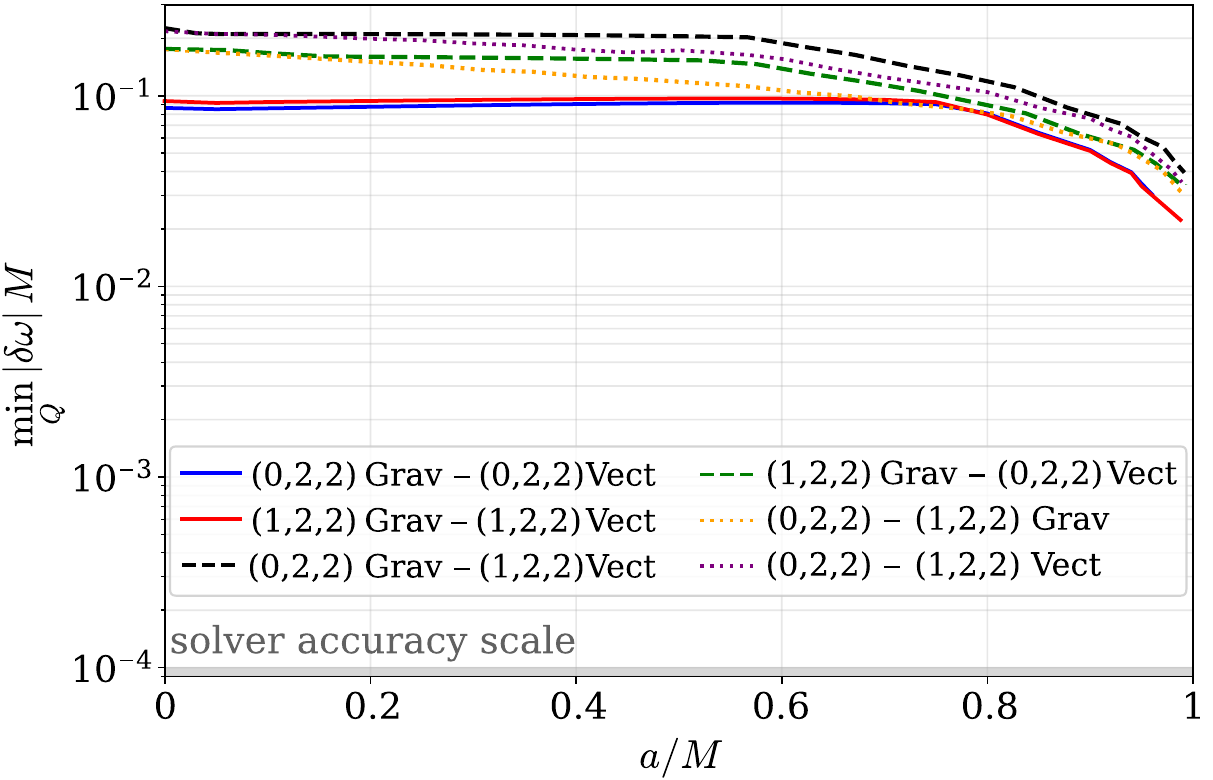}\\
        \caption{(Left) schematic representation of the eigenvalue repulsion behaviour between two mode families as a function of the coupling parameter ($p$). (Right) Closest approach between the Grav- and Vect-led branches of the $m=2$ PS modes, quantified by $\min_Q | \delta \omega| M$ as a function of spin. Line style groups the pairing: same-mode inter-class (solid); cross-mode inter-class (dashed); overtone intra-class (dotted).}
        \label{F4}
    \end{figure}

    The mathematical framework for such avoided crossings has been developed by Yang, Berti and Franchini \cite{yang2025black}, who showed that in systems with two or more free parameters, avoided crossings are generically associated with exceptional points -- isolated degeneracies where both frequencies and eigenfunctions coalesce, and the local frequency splitting follows a characteristic $\sqrt{p -\hat{p}}$ topology (see Fig.~\ref{F4}, left).

    The present dataset enables:
    \begin{itemize}
        \item[\textit{(i)}] tracking the onset of the eigenvalue repulsion between the Grav-led PS $(1,2,2)$ and the $(0,2,2)$ NH families of solutions;
        \item[\textit{(ii)}] assess whether such repulsion is also present in the Vect-led class;
        \item[\textit{(iii)}] study whether the two polarizations of the same PS ($n, \ell, m$) interact -- a type of repulsion that exists only in coupled multi-field systems and has not been investigated.
    \end{itemize}
    Addressing point \textit{(iii)} first, across the surveyed $(a,\, Q)$ domain, the Grav- and Vect-led branches of the $m=2$ PS mode exhibit no coincidence in either the real or imaginary part, and their complex separation $|\delta \omega|$ is monotonic with no local minimum (see Fig.~\ref{F4}); by the argument of \cite{yang2025black}, this excludes an exceptional point -- and with it inter-class eigenvalue repulsion -- within the double-precision-resolved region.
    Moving to the eigenvalue repulsion between the PS $(1,2,2)$ and $(0,2,2)$ NH families reported by Dias \textit{et al.}~\cite{dias2022eigenvalue}, we must now point out that both the PS and NH eigenfunctions become poorly resolved near the eigenvalue repulsion boundary identified in~\cite{dias2022eigenvalue}, where the electromagnetic spectral component is suppressed below the \texttt{float64} precision floor. A definitive statement on the repulsions at high charge therefore requires quadruple-precision arithmetic in the linear solve, which we defer to future work. With our current dataset, however, we can try to detect the onset of the eigenvalue repulsion on the PS $(1,2,2)$\footnote{We would like to point out that, while eigenvalue repulsion is expected to be observed also between the PS $(0,2,2)$ and the NH $(1,2,2)$, such falls outside our regime. The PS $(1,2,2)$ is the only one that presents the repulsion inside our region of trust.}.

\begin{figure}[h!]
        \centering
        \includegraphics[width=0.85\textwidth]{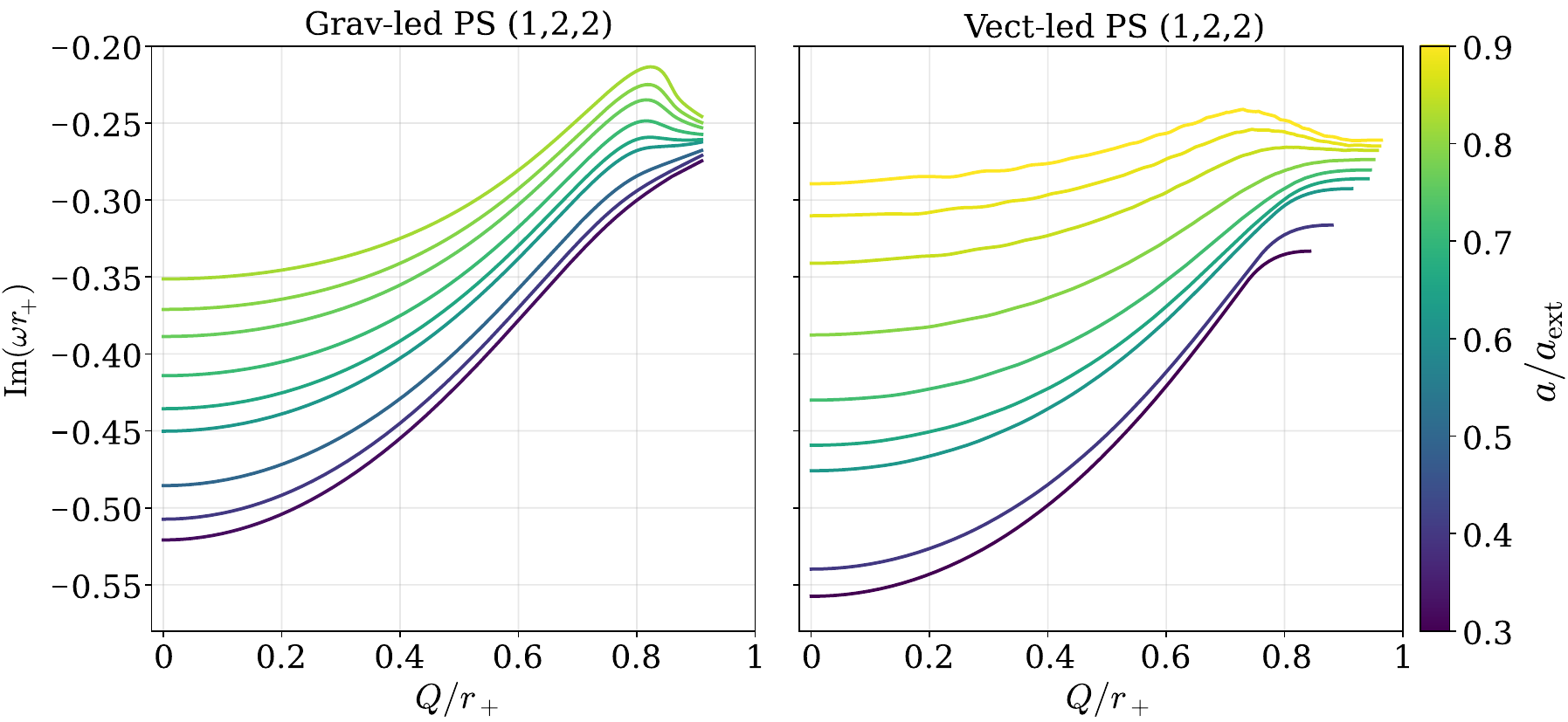}\\
        \caption{Im$(\omega r_+)$ of the PS $(1,2,2)$ mode vs $Q/r_+$ along constant-$a/a_{\rm ext}$ slices, for the Grav-led (left) and Vect-led (right) branches. The Grav-led curves turn over near $Q/r_+\approx0.82$ -- the onset of the repulsion of \cite{dias2022eigenvalue}; the Vect-led curves turn over only at $a/a_{\rm ext}\approx0.90$, in the near-extremal region ($Q/Q_{\rm ext}\gtrsim0.97$) where onset and numerical breakdown are indistinguishable.}
        \label{F5}
    \end{figure}

    For direct comparison with \cite{dias2022eigenvalue}, we report this analysis in horizon units, $\omega r_+$, rather than the mass-normalized $\omega M$ used above. In Fig.~\ref{F5} (left) we plot the Im$(\omega r_+)$ component of the PS $(1,2,2)$ dataset as a function of $Q/r_+$, sampled across constant $a/a_{\rm ext}$ slices. While for small $a/a_{\rm ext}$ slices the Im$(\omega r_+)$ line is monotonically increasing, for $a/a_{\rm ext}\gtrsim 0.62$, the Im$(\omega r_+)$ curve acquires a local maximum at $Q/r_+\approx0.82$, then turns over. This is the onset of the avoided crossing reported by Dias \textit{et al.}~\cite{dias2022eigenvalue}, observed here from the PS side -- point \textit{(i)}. The lower branch, however, is not recovered since the solver lacks the accuracy required to resolve the gap and when performing continuation from $q=0$ it rides the upper sheet. The near-degeneracy sits where $\nabla \omega$ is singular~\cite{yang2025black}, so the NR continuation must fail there -- resolving the gap requires extended-precision seeding onto the lower sheet.
    Applying now the same analysis to point \textit{(ii)}, Fig.~\ref{F5} (right) shows the constant $a/a_{\rm ext}$ slices for Vect-led PS $(1,2,2)$ mode. For $a/a_{\rm ext}<0.85$ the Im$(\omega r_+)$ lines are monotonically increasing through the reliable range, and hence no onset of repulsion where the Grav-led one occurs. The onset signature is, therefore, polarization dependent. For $a/a_{\rm ext}$ between $0.88 - 0.90$, a similar bend on the Im$(\omega r_+)$ line emerges near the extremal limit ($Q/Q_{ext}\gtrsim 0.97)$ where the onset and the numerical breakdown are indistinguishable. Observe that the Grav-led bend is driven by the approaching Grav-led NH $(0,2,2)$; the Vect-led analogue would need a Vect-led NH partner, which the solver cannot resolve -- seed collapses to Grav-led. Full confirmation of the repulsion requires extended precision and will also be left for future work.

%
    \subsection{Einstein Telescope forecast}\label{S43}
%
    The QNM frequencies computed in this work encode how the ringdown signal of a KN BH departs from the Kerr prediction as a function of the charge, $Q$. In this section, we shall translate this spectral information into projected measurement uncertainties on $Q/M$ using the Fisher matrix formalism, adopting the Einstein Telescope in its ET-D configuration as the reference detector.

    Following the framework established in \cite{pombo2026teukolsky}, let us model the ringdown contribution of each QNM, labeled $k=(n,\ell,m)$, as a linearly polarized damped sinusoid,
    \begin{equation}
        h_k (t) = A_k e^{-\frac{t}{\tau_k}}\cos \big( 2\pi f_k\ t+\phi _k\big)\ ,
    \end{equation}
    defined for $t\geqslant 0$ and zero otherwise. The physical frequency and damping time are obtained from the dimensionless QNM frequency $\hat{\omega}_k (a/M, Q/M)$ by fixing a source-frame remnant mass $M$,
    \begin{equation}
        f_k = \frac{{\rm Re} (\hat{\omega}_k) c^3}{2\pi G M }\ ,\qquad \qquad \tau_k = \frac{GM}{|{\rm Im}(\hat{\omega}_k)|c^3}\ ,
    \end{equation}
    so that all dependence on the BH parameters, $\Theta =\{M,\, a/M,\, Q/M\}$, enters through $\hat{\omega}_k (\Theta)$. For a multi-mode ringdown, the total signal is $h(t)=\Sigma_k\, h_k (t)$, where the sum runs over the detected modes. The ET detector is modeled as a single effective detector characterized by the sum amplitude spectral density, $S_h ^{\rm sum} (f)$. The noise-weighted inner product between two signals is
    \begin{equation}
        \big( g_1 |g_2 \big) = 4\, {\rm Re}\Bigg( \int_{f_{\min}} ^{f_{\max}}\frac{\tilde{g}_1 (f)\, \tilde{g}_2 ^*(f)}{S_n (f)}\, \dd f\Bigg)\ , 
    \end{equation}
    with $S_n (f) \equiv S_h ^{\rm sum} (f)$, integration band $2\, {\rm Hz}\leqslant f\leqslant 2048\, {\rm Hz}$ and optimal SNR $\rho_k ^2 = \big(h_k |h_k\big)$.

    For a signal that depends on parameters $\Theta = \{ \Theta _i\}$, the Fisher information matrix is
    \begin{equation}
        \Gamma _{ij}=\bigg(\frac{\partial h}{\partial \Theta _i}\bigg|\frac{\partial h}{\partial \Theta _j}\bigg)\ .
    \end{equation}
    In the high-SNR limit, the covariance of any unbiased estimator satisfies Cov$(\Theta_i, \Theta _j) \geqslant (\Gamma _{ij})^{-1}$, so the marginalized $1\sigma$ uncertainty on a single parameter is $\sigma (\Theta _i)=\sqrt{(\Gamma^{-1})_{ii}}$. 

    Via the chain rule, the waveform derivatives reduce to derivatives of the QNM frequency:
    \begin{equation}
        \frac{\partial f_k}{\partial \Theta_i} = \frac{c^3}{2\pi GM}\frac{\partial {\rm Re}(\hat{\omega}_k)}{\partial \Theta _i}\ , \qquad \qquad \frac{\partial \tau_k}{\partial \Theta _i} = -\frac{GM}{c^3\big[{\rm Im}(\hat{\omega}_k)\big]^2} \frac{\partial  {\rm Im}(\hat{\omega}_k)}{\partial \Theta_i}\ .
    \end{equation}
    The charge susceptibilities $\partial \hat{\omega}_k /\partial(Q/M)^2$ -- the key quantities connecting the QNM dataset to the observable constraints -- are computed numerically from the dense charge sampling along each spin slice. 
    
    For a multi-mode detection, the total Fisher matrix is the sum of the per-mode contributions, $\Gamma^{\rm Total}_{ij}=\sum_k\Gamma^k_{ij}$, since the well-separated modes are approximately orthogonal in the noise-weighted inner product. For a damped sinusoid in the high-SNR limit, the Fisher matrix block-diagonalizes \cite{Berti2006}: the per-mode amplitudes, $A_k$, and phases, $\phi_k$, decouple from the frequencies and damping times. We therefore construct the Fisher matrix directly in the three-dimensional intrinsic parameter space. Due to the fact that the KN spectrum depends on the charge solely through $Q^2$, the linear response $\partial\hat{\omega}/\partial(Q/M)$ vanishes identically at $Q/M=0$. The natural Fisher parameter is therefore $(Q/M)^2$, for which the spectrum has a non-vanishing linear response. The Fisher matrix is constructed in $\Theta=\big\{M,\, a/M,\, (Q/M)^2\big\}$, and the charge bound is reported as $(Q/M)_{\rm bound}=\big[\sigma_{(Q/M)^2}\big]^{1/2}$, which scales as $\rho^{-1/2}$ rather than the $\rho^{-1}$ characteristic of parameters to which the waveform responds linearly.

\begin{figure}[h!]
        \centering
        \includegraphics[width=0.9\textwidth]{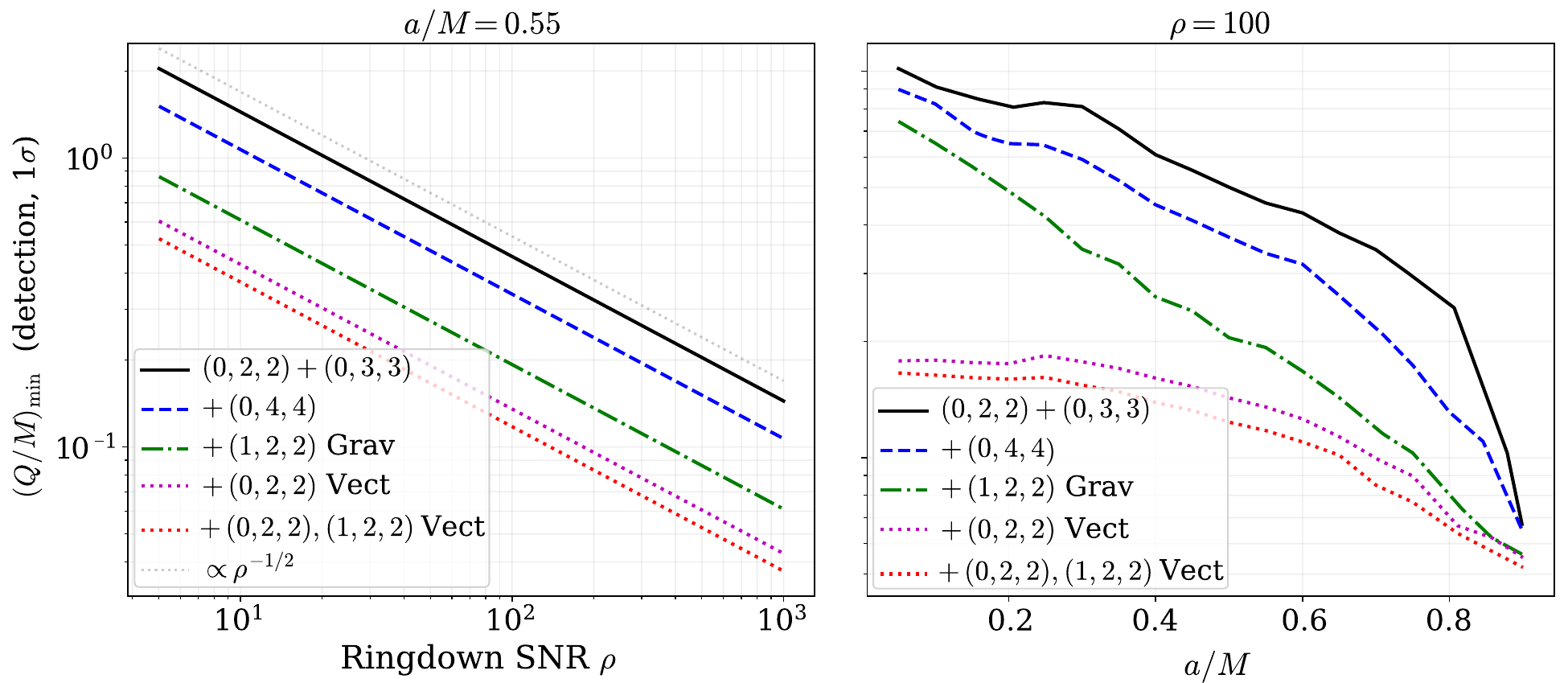}\\
        \caption{Fisher-matrix charge forecast for a $M=70\,M_\odot$ KN remnant detected by ET-D, with all scenarios marginalizing over $\big\{M,\, a/M,\, (Q/M)^2\big\}$. (Left) detection threshold $(Q/M)_{\rm bound}$ versus ringdown SNR $\rho$ at fiducial spin $a/M=0.55$. (Right) detection threshold versus $a/M$ at fixed $\rho=100$, for a two-mode baseline $(0,2,2)+(0,3,3)$ (solid black); a three-mode combination adding $(0,4,4)$ (dashed blue); a four-mode combination further adding the $(1,2,2)$ overtone (dot-dashed green); the four PS Grav-led modes supplemented with the $(0,2,2)$ PS Vect-led mode (dotted purple), and both the $(0,2,2)$ and $(1,2,2)$ Vect-led modes (dotted red).}
        \label{F6}
    \end{figure}

    A single QNM provides two observables ($f_k$ and $\tau_k$), which is insufficient to constrain the three intrinsic parameters simultaneously; at least two modes are required for a non-degenerate Fisher matrix. For this exercise, the considered fiducial remnant has source-frame mass $M=70\,M_\odot$ and is placed at a luminosity distance $d_L$ such that the dominant PS mode $(0,2,2)$ has a specified ringdown SNR $\rho$. For the detection forecast, the Kerr limit serves as the reference point; the Fisher matrix is evaluated at $Q/M=0$ to obtain the projected sensitivities on charge under the null hypothesis -- that is, the $1\sigma$-equivalent charge scale that could be distinguished from Kerr at a given SNR. We stress that, because $Q/M=0$ lies on the boundary of the physical domain $(Q/M)^2 \geq 0$, the Gaussian-linear Fisher estimate is a forecast scale rather than a strict frequentist threshold; a rigorous near-null bound would require a Bayesian or upper-limit treatment, which we defer to a dedicated data-analysis study. Assuming each mode in a given combination is detected at the same ringdown SNR $\rho$, we present two complementary forecasts:
    
\vspace{1mm}

    \textbf{Low-charge sensitivity:} In the first, more realistic scenario, let us compare multi-mode PS Grav-led combinations of increasing richness, namely: a two-mode baseline $(0,2,2)+(0,3,3)$ (solid black), representative of the minimum information from a two-tone ringdown detection; a three-mode combination adding $(0,4,4)$ (dashed blue); and a four-mode combination further adding the $(1,2,2)$ overtone (dot-dashed green). All scenarios marginalize jointly over $\big\{M,\, a/M,\, (Q/M)^2\big\}$, and the bound scales as $(Q/M)_{\rm bound}\propto\rho^{-1/2}$ (dotted grey) in each case (Fig.~\ref{F6}, left). At a reference ringdown SNR of $\rho=100$ and fiducial spin $a/M=0.55$ (Fig.~\ref{F6}, right), the two-mode baseline yields $(Q/M)_{\rm bound}\approx 0.46$, tightening to $\approx 0.34$ with the addition of $(0,4,4)$ ($\approx 1.35\times$) and to $\approx 0.19$ once the $(1,2,2)$ overtone is included ($\sim 1.8\times$ relative to three modes). The overtone is the single most informative addition: although it shares the angular structure of the dominant $(0,2,2)$ mode, its markedly shorter damping time breaks parameter degeneracies that the angular harmonics alone do not. The bound tightens further toward higher spin, where the charge susceptibilities grow, and all multi-mode combinations maintain consistent ordering across the spin range.

\vspace{1mm}

\begin{figure}[h!]
        \centering
        \includegraphics[width=1\textwidth]{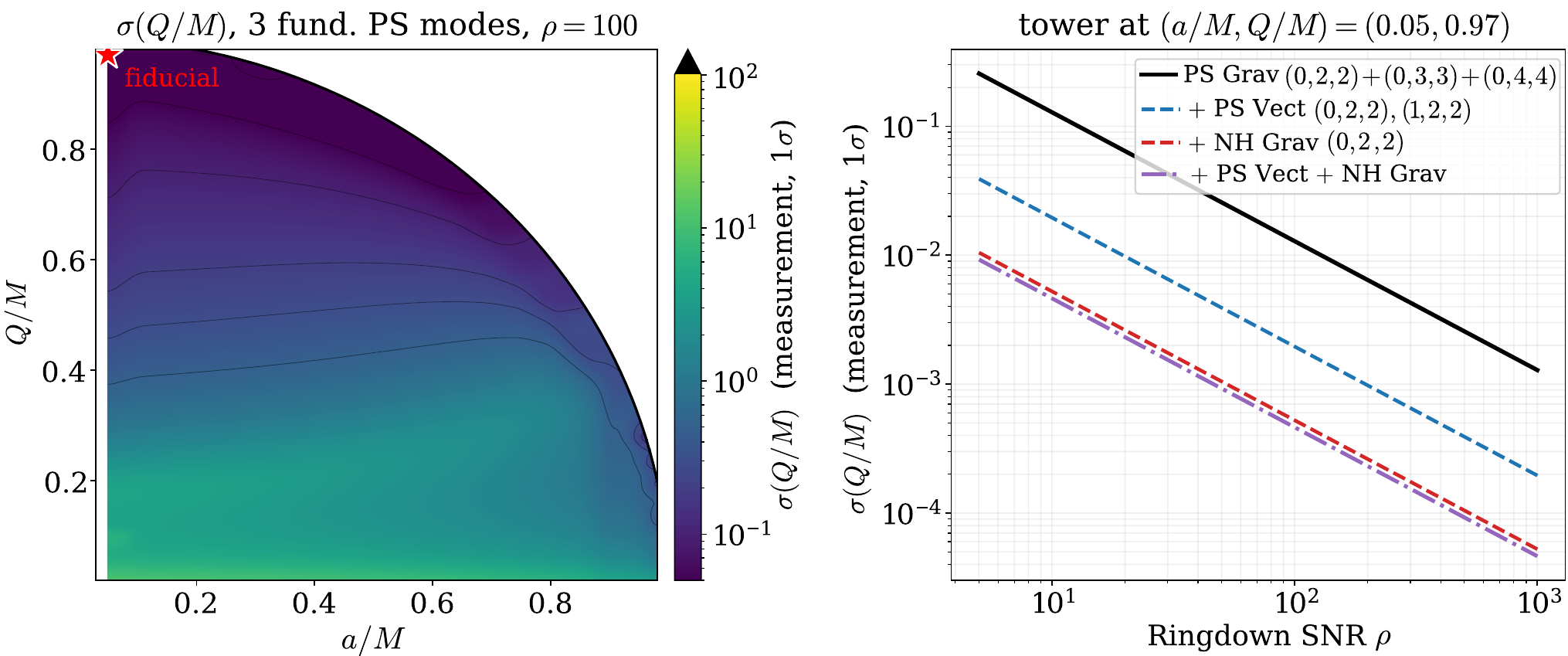}\\
        \caption{Fisher-matrix charge forecast for a $M=70\,M_\odot$ KN remnant detected by ET-D, with all scenarios marginalizing over $\big\{M,\, a/M,\, (Q/M)^2\big\}$. (Left) measurement uncertainty $\sigma(Q/M)$ over the $(a/M,\,Q/M)$ plane for the three fundamental PS Grav-led modes at $\rho=100$; the colour scale is logarithmic, the solid curve marks the extremality bound $a^2+Q^2=1$, and the starred marker indicates the near-extremal fiducial $(a/M,\,Q/M)=(0.05,\,0.97)$ used in the right panel. (Right) measurement uncertainty $\sigma(Q/M)$ versus ringdown SNR $\rho$ at the fiducial point, for the three PS Grav-led fundamentals $(0,2,2)+(0,3,3)+(0,4,4)$ (solid black); supplemented with the PS Vect-led $(0,2,2)$ and $(1,2,2)$ modes (dashed blue); with the NH Grav-led $(0,2,2)$ mode (dashed red); and with both PS Vect-led and NH Grav-led additions combined (dash-dotted purple).}
        \label{F7}
    \end{figure}

    \textbf{The Grav--Vect PS lever arm:} As an illustrative exercise, we assess the spectroscopic information carried by the Vect-led branch, setting aside its excitation. The four PS Grav-led modes are supplemented first with the $(0,2,2)$ PS Vect-led mode (dotted purple), and then with both the $(0,2,2)$ and $(1,2,2)$ Vect-led modes (dotted red), assuming the latter are excited to a comparable ringdown SNR. We stress that this is not a realistic detection scenario -- the Vect-led gravitational excitation is $Q^2$-suppressed (Sec.~\ref{S41}) -- but it quantifies the constraining power these modes would contribute if such suppression were overcome. Because the Grav-led and Vect-led branches of the same $(n,\ell,m)$ respond to charge with distinct susceptibilities, their combination provides a lever arm in $Q/M$ that a single-polarization analysis lacks. At $\rho=100$ (Fig.~\ref{F6}, right), adding the $(0,2,2)$ Vect-led mode tightens the four-mode Grav-led bound from $\approx 0.19$ to $\approx 0.14$ (a factor of $\sim 1.4$), with the further inclusion of the $(1,2,2)$ Vect-led mode yielding only a marginal additional gain (to $\approx 0.12$). The Vect-led contribution is thus genuine but modest: the distinct charge response of the Vect-led branch adds independent information, but does not dramatically outperform the Grav-led tower on its own.   
    Complementing the detection forecasts above, which are anchored at $Q/M=0$, we also assess the measurement precision for a remnant that carries a nonzero charge. For each point in the $(a/M,\,Q/M)$ plane we evaluate the three-fundamental-mode Fisher matrix at that fiducial value and extract the marginalized uncertainty $\sigma(Q/M)$ via standard error propagation from $\sigma_{(Q/M)^2}$. For this study, besides the PS Grav- and Vect-led branches, one can consider also the NH family of solutions:

    \vspace{1mm}
    \textbf{The PS--NH Grav-led lever arm:} The NH family admits no $Q/M=0$ limit and so cannot enter the Kerr-anchored detection forecast; we instead assess it within the finite-charge measurement question above, at a near-extremal-charge fiducial $(a/M,\,Q/M)=(0.05,\,0.97)$ where the NH $(0,2,2)$ mode is well resolved and in-band $(\approx60\,$Hz for $M=70\,M_\odot)$. There, the PS and NH $(0,2,2)$ modes are spectroscopically well separated and respond to charge very differently -- the PS damping is nearly charge-independent while the NH damping varies roughly two orders of magnitude more steeply -- so their combination supplies a charge lever arm considerably stronger than the Vect--Grav pairing. The spatial dependence of $\sigma(Q/M)$ across the sub-extremal parameter space is shown in Fig.~\ref{F7} (left), where the fiducial point is marked. At $\rho=100$ (Fig.~\ref{F7}, right), the three PS Grav-led fundamentals alone yield $\sigma(Q/M)\approx1.3\times10^{-2}$. Supplementing them with the PS Vect-led $(0,2,2)$ and $(1,2,2)$ modes tightens this to $\approx2.0\times10^{-3}$, while adding instead the single NH Grav-led $(0,2,2)$ mode gives $\approx5.2\times10^{-4}$ -- a factor $\sim25$ improvement over the PS-only bound, reflecting the steep near-extremal charge response of the NH damping. Combining all three additions yields only a marginal further gain, to $\approx4.6\times10^{-4}$, confirming that the NH mode supplies the dominant charge leverage at this fiducial. This is a deliberately illustrative, near-extremal-charge estimate: the susceptibilities are finite-differenced from fixed-charge NH slices, and the steep near-extremal response together with the \texttt{float64} precision wall (Sec.~\ref{S42}) makes the bound indicative rather than a production forecast.
    \vspace{1mm}

    It is also important to point out that while the $(0,2,0)$ NH modes can exhibit strong charge dependence, their real frequencies remain very small across the parameter space (${\rm Re}(\hat{\omega})\lesssim 5\times10^{-3}$), placing them below $\sim 1\,{\rm Hz}$ for a $70\,M_\odot$ remnant -- well beneath the ET low-frequency cutoff of $2\,{\rm Hz}$. The NH family therefore falls outside the detector band for stellar-mass remnants, regardless of its charge sensitivity or damping properties.
    
    Finally, while astrophysical BHs are expected to carry negligible charge, these bounds apply more broadly. Beyond-GR theory whose perturbations form a coupled two-field system with a mixing parameter can in principle be mapped onto the KN framework by identifying that parameter with an effective $Q/M$. In this way, observational bounds on $Q/M$ translate into a bound on the coupling of the theory in question.
    
%
\section{Conclusion}\label{S5}
%
    In this work, we extend \texttt{SpectralPINN}, a hybrid pseudo-spectral/physics informed neural network (PINN) solver, to solve the coupled PDE master system describing the quasinormal modes of a Kerr-Newman black hole. The improved solver was benchmarked against the Kerr $(Q=0)$ regime \cite{berti2009quasinormal,berti2009ringdown}, the Reissner--Nordstr\"om $(a=0)$ regime \cite{leaver1990quasinormal,kokkotas1988black,matyjasek2017quasinormal}, and the Kerr--Newman $a=Q$ diagonal \cite{dias2015linear,berti2009ringdown}, yielding a maximum relative combined $(\mathbb{R}+\mathbb{I})$ error $\sim 10^{-4}$ for the $(1,2,2)$ mode (worst case). The remaining computed modes possess relative combined errors $\sim 10^{-7}$, with the minimum reaching $10^{-9}$. This represents a three-orders-of-magnitude improvement compared to the previous version \cite{pombo2026teukolsky}. The increase in accuracy comes prominently through the adoption of a second-order optimizer based on a NR scheme, which, in conjunction with a ``warm-up'' of the solution through the previously developed HybridAdamD solver, creates a two-stage optimization able to reach residual norms $\sim 10^{-12}$ (corresponding to the $10^{-7}$ relative error in the frequency).
    
    To illustrate the computational power of \texttt{SpectralPINN}, we computed the quasinormal frequencies across a wide range of $(n,\ell,m)$ modes, effectively mapping their frequency dependence with the $(M,\, a/M,\, Q/M)$ parameters for the photon-sphere $(0,2,2)$, $(0,2,0)$, $(0,3,3)$, $(0,4,4)$ and $(1,2,2)$ both Grav- and Vect-led modes, and the near-horizon $(0,2,0)$ and $(0,2,2)$ Grav-led modes. This extensive dataset is publicly released in a Zenodo repository together with the generating equations \cite{Zenodo}.

    Operating in double precision, the Newton-Raphson continuation breaks down precisely at the photon-sphere/near-horizon eigenvalue-repulsion boundary~\cite{dias2022eigenvalue}, where the near-horizon eigenfunction develops support away from the horizon and the linear system becomes severely ill-conditioned. This is intrinsic to the near-degeneracy rather than a numerical accident: the parameter gradient, $\nabla \omega$, is singular at the exceptional point (Sec.~\ref{S42}), so the continuation necessarily fails as it is approached. Resolving the frequency gap across the repulsion therefore requires extended-precision arithmetic in the linear solve, which we leave to future work.

    Even within this limitation, the resolved portion of the spectrum already exhibits the signatures of the repulsion. Tracking the Grav-led PS $(1,2,2)$ branch along constant-$a/a_{\rm ext}$ slices, it is possible to observe a damping turn over as it approaches the near-horizon $(0,2,2)$ family -- the onset of the avoided crossing reported in~\cite{dias2022eigenvalue}, seen here from the photon-sphere side, even though the gap itself lies beyond our reach. Applying the exceptional-point criterion of~\cite{yang2025black}, we further find no coincidence between the Grav- and Vect-led branches of any $m=2$ photon-sphere mode across the resolved domain, excluding an inter-class repulsion between the two polarizations. A similar onset signature appears in the Vect-led sector near extremality ($a/a_{\rm ext}\approx0.9$); it falls within the near-extremal regime where physical onset and numerical breakdown cannot yet be separated, and its confirmation must await the extended-precision treatment noted above.

    With the computed dataset, we proceeded to map the change of the frequency with the charge by computing the Fisher matrix for a combination of multi-mode, multi-branch configurations. The charge susceptibilities computed from the dense charge sampling -- the derivative in $(Q/M)^2$ at the Kerr limit $Q\to0$ -- feed a Fisher matrix in $\big\{M,a/M,(Q/M)^2\big\}$. The marginalized projected detection thresholds scale as $\rho^{-1/2}$. In a single-branch analysis, the multi-mode improvement with the $(1,2,2)$ mode proves to be the single most informative addition ($0.34 \to 0.19$ at $\rho=100$). In the illustrative equal-SNR scenario, including the Vect-led branch further tightens the charge bound $(Q/M)_{\rm bound}$ from $0.19\to 0.14$, though with a lower impact than the Grav-led first overtone. 

    Beyond the analyses presented here, the released dataset \cite{Zenodo} provides the raw coupled frequencies across the sub-extremal plane, where previously only the $a = Q$ diagonal of the PS $(n,\ell,m) = (0,2,2)$ Grav-led mode was publicly available. It can serve to calibrate separable approximations such as Dudley--Finley~\cite{dudley1977separation,dudley1979covariant,saha2025revisiting} -- now extendable to the Vect-led sector, for which no exact counterpart previously existed --, to supply exact Kerr-Newman frequencies for ringdown inference pipelines, and to provide the densely sampled two-parameter spectra required by the exceptional-point framework.
    
    Finally, we would like to point out that, while astrophysical black holes are expected to carry negligible electric charge, the Kerr--Newman solution, due to the existence of a system of master equations, constitutes an important step towards computing quasinormal modes of alternative compact objects, as it contains an additional field (electromagnetic) coupled to gravity and works as a toy model to study phenomenology (\textit{e.g.,} eigenvalue repulsion, resonances) that are absent in the standard Kerr solution.
%
    \section*{Data availability} \label{Akn}
%
    The complete dataset -- complex frequencies $\omega (a/M, Q/M)$ for all $12$ mode branches is publicly released at \cite{Zenodo}. We also provide the explicit factorization coefficient functions entering the coupled PDE system \eqref{E2.16}--\eqref{E2.19} in both \texttt{Mathematica} and \texttt{PyTorch} format. Additional modes and/or specific mode amplitudes (eigenfunctions) can be provided upon request.

%
\acknowledgments
    We would like to thank Ignacy Sawicki, Carlos Herdeiro and Lorenzo Pizzuti for their valuable discussions and comments. A. M. Pombo is supported by the Czech Grant Agency (GA\^CR) project PreCOG (Grant No. 24-10780S). We acknowledge the support of the European Consortium for Astroparticle Theory in the form of an Exchange Travel Grant. 
%

\bibliographystyle{JHEP}
\bibliography{biblio}

@article{Blazquez2024,
  author       = {Bl{\'a}zquez-Salcedo, J.~L. and Khoo, F.~S. and Kunz, J. and Gonz{\'a}lez-Romero, L.~M.},
  title        = {Quasinormal modes of Kerr black holes using a spectral decomposition of the metric perturbations},
  journal      = {Phys.\ Rev.\ D},
  volume       = {109},
  number       = {6},
  pages        = {064028},
  year         = {2024},
  doi          = {10.1103/PhysRevD.109.064028}
}

@article{LVK:2025_GW250114_KerrArea,
    author        = "Abac, A. G. and others",
    collaboration = "LIGO Scientific Collaboration, Virgo Collaboration and KAGRA Collaboration",
    title         = "{GW250114: Testing Hawking's Area Law and the Kerr Nature of Black Holes}",
    journal       = "Phys. Rev. Lett.",
    volume        = "135",
    number        = "11",
    pages         = "111403",
    year          = "2025",
    doi           = "10.1103/kw5g-d732",
    eprint        = "2509.08054",
    archivePrefix = "arXiv",
    primaryClass  = "gr-qc"
}

@article{LVK:2026_GW250114_Spectroscopy,
    author        = "Abac, A. G. and others",
    collaboration = "LIGO Scientific Collaboration, Virgo Collaboration and KAGRA Collaboration",
    title         = "{Black Hole Spectroscopy and Tests of General Relativity with GW250114}",
    journal       = "Phys. Rev. Lett.",
    volume        = "136",
    number        = "4",
    pages         = "041403",
    year          = "2026",
    doi           = "10.1103/6c61-fm1n",
    eprint        = "2509.08099",
    archivePrefix = "arXiv",
    primaryClass  = "gr-qc"
}

@misc{berti2009ringdown,
  author = {Berti, E.},
  title  = {Ringdown},
  year   = {2009},
  url    = {https://pages.jh.edu/eberti2/ringdown/},
  note   = {Online data tables for Kerr QNM frequencies},
}

@article{berti2009quasinormal,
  author  = {Berti, E. and Cardoso, V. and Starinets, A. O.},
  title   = {Quasinormal modes of black holes and black branes},
  journal = {Class. Quant. Grav.},
  volume  = {26},
  pages   = {163001},
  year    = {2009},
  eprint  = {0905.2975},
}

@article{pombo2026teukolsky,
  title={Teukolsky by design: A hybrid spectral-PINN solver for Kerr quasinormal modes},
  author={Pombo, Alexandre M and Pizzuti, Lorenzo},
  journal={Journal of Cosmology and Astroparticle Physics},
  volume={2026},
  number={03},
  pages={009},
  year={2026},
  publisher={IOP Publishing}
}

@article{GWTC1,
  author       = {Abbott, B.~P. et al. (LIGO Scientific Collaboration and Virgo Collaboration)},
  title        = {GWTC-1: A Gravitational-Wave Transient Catalog of Compact Binary Mergers Observed by LIGO and Virgo during the First and Second Observing Runs},
  journal      = {Phys. Rev. X},
  volume       = {9},
  number       = {3},
  pages        = {031040},
  year         = {2019},
  doi          = {10.1103/PhysRevX.9.031040}
}

@article{GWTC2,
  author       = {Abbott, R. et al. (LIGO Scientific Collaboration and Virgo Collaboration)},
  title        = {GWTC-2: Compact Binary Coalescences Observed by LIGO and Virgo during the First Half of the Third Observing Run},
  journal      = {Phys. Rev. X},
  volume       = {11},
  number       = {2},
  pages        = {021053},
  year         = {2021},
  doi          = {10.1103/PhysRevX.11.021053}
}

@article{GWTC3,
  author       = {Abbott, R. et al. (LIGO Scientific, Virgo and KAGRA Collaborations)},
  title        = {GWTC-3: Compact Binary Coalescences Observed by LIGO and Virgo during the Second Half of the Third Observing Run},
  journal      = {arXiv preprint},
  eprint       = {2111.03606},
  year         = {2021},
  url          = {https://arxiv.org/abs/2111.03606}
}

@article{LVK_open_data_O4a,
  author         = "{The LIGO Scientific Collaboration and the Virgo Collaboration and the KAGRA Collaboration}",
  title          = "{Open data from LIGO, Virgo, and KAGRA through the first half of the fourth observing run}",
  eprint         = "2508.18079",
  archivePrefix  = "arXiv",
  primaryClass   = "astro-ph.HE",
  year           = "2025"
}

@article{Punturo2010ET,
  author        = {Punturo, M. and others},
  title         = {The Einstein Telescope: A third-generation gravitational wave observatory},
  journal       = {Class. Quant. Grav.},
  volume        = {27},
  pages         = {194002},
  year          = {2010},
  doi           = {10.1088/0264-9381/27/19/194002},
  eprint        = {1002.0462},
  archivePrefix = {arXiv},
  primaryClass  = {gr-qc}
}

@article{Evans2021CE,
  author        = {Evans, M. and others},
  title         = {A Horizon Study for Cosmic Explorer: Science, Observatories, and Community},
  journal       = {arXiv e-prints},
  year          = {2021},
  eprint        = {2109.09882},
  archivePrefix = {arXiv},
  primaryClass  = {gr-qc}
}

@article{Hall2022CE,
  author        = {Hall, E. D. and others},
  title         = {Cosmic Explorer: A Next-Generation Ground-Based Gravitational-Wave Observatory},
  journal       = {Galaxies},
  volume        = {10},
  pages         = {90},
  year          = {2022},
  doi           = {10.3390/galaxies10040090}
}

@article{dias2015linear,
  title={Linear mode stability of the Kerr-Newman black hole and its quasinormal modes},
  author={Dias, Oscar JC and Godazgar, Mahdi and Santos, Jorge E},
  journal={Physical review letters},
  volume={114},
  number={15},
  pages={151101},
  year={2015},
  publisher={APS}
}

@article{yang2013quasinormal,
  title={Quasinormal modes of nearly extremal Kerr spacetimes:<? format?> Spectrum bifurcation and power-law ringdown},
  author={Yang, Huan and Zimmerman, Aaron and Zengino{\u{g}}lu, An{\i}l and Zhang, Fan and Berti, Emanuele and Chen, Yanbei},
  journal={Physical Review D—Particles, Fields, Gravitation, and Cosmology},
  volume={88},
  number={4},
  pages={044047},
  year={2013},
  publisher={APS}
}

@article{yang2013branching,
  title={Branching of quasinormal modes for nearly extremal Kerr black holes},
  author={Yang, Huan and Zhang, Fan and Zimmerman, Aaron and Nichols, David A and Berti, Emanuele and Chen, Yanbei},
  journal={Physical Review D—Particles, Fields, Gravitation, and Cosmology},
  volume={87},
  number={4},
  pages={041502},
  year={2013},
  publisher={APS}
}

@article{zimmerman2016damped,
  title={Damped and zero-damped quasinormal modes of charged, nearly extremal black holes},
  author={Zimmerman, Aaron and Mark, Zachary},
  journal={Physical Review D},
  volume={93},
  number={4},
  pages={044033},
  year={2016},
  publisher={APS}
}

@article{matyjasek2017quasinormal,
  title={Quasinormal modes of black holes: The improved semianalytic approach},
  author={Matyjasek, Jerzy and Opala, Micha{\l}},
  journal={Physical Review D},
  volume={96},
  number={2},
  pages={024011},
  year={2017},
  publisher={APS}
}

@article{dias2022eigenvaluef,
  title={Eigenvalue repulsions and quasinormal mode spectra of Kerr-Newman: an extended study},
  author={Dias, Oscar JC and Godazgar, Mahdi and Santos, Jorge E},
  journal={arXiv preprint arXiv:2205.13072},
  year={2022}
}

@article{dias2022eigenvalue,
  title={Eigenvalue repulsions in the quasinormal spectra of the Kerr-Newman black hole},
  author={Dias, Oscar JC and Godazgar, Mahdi and Santos, Jorge E and Carullo, Gregorio and Del Pozzo, Walter and Laghi, Danny},
  journal={Physical Review D},
  volume={105},
  number={8},
  pages={084044},
  year={2022},
  publisher={APS}
}

@article{mark2015quasinormal,
  title={Quasinormal modes of weakly charged Kerr-Newman spacetimes},
  author={Mark, Zachary and Yang, Huan and Zimmerman, Aaron and Chen, Yanbei},
  journal={Physical Review D},
  volume={91},
  number={4},
  pages={044025},
  year={2015},
  publisher={APS}
}

@article{LISA2017,
  author        = {Amaro-Seoane, Pau and Audley, Heather and Babak, Stanislav and Baker, John and Barausse, Enrico and Bender, Peter and et al.},
  title         = {Laser Interferometer Space Antenna},
  journal       = {arXiv e-prints},
  eprint        = {1702.00786},
  archivePrefix = {arXiv},
  primaryClass  = {astro-ph.IM},
  year          = {2017}
}

@article{ishak2018gravitational,
  title={Gravitational waves volume 2: astrophysics and cosmology},
  author={Ishak, B},
  journal={Contemporary Physics},
  year={2018},
  publisher={Taylor \& Francis}
}

@article{kokkotas1999quasi,
  title={Quasi-normal modes of stars and black holes},
  author={Kokkotas, Kostas D and Schmidt, Bernd G},
  journal={Living Reviews in Relativity},
  volume={2},
  number={1},
  pages={2},
  year={1999},
  publisher={Springer}
}

@article{Isi2021,
  author       = {Isi, M. and Farr, W.~M. and Chatziioannou, K. and Giesler, M. and Hughes, S.~A.},
  title        = {Testing the black-hole no-hair theorem with gravitational waves},
  journal      = {Phys. Rev. Lett.},
  volume       = {127},
  number       = {1},
  pages        = {011103},
  year         = {2021},
  doi          = {10.1103/PhysRevLett.127.011103}
}

@article{Berti2006,
  author       = {Berti, E. and Cardoso, V. and Will, C.~M.},
  title        = {Gravitational-wave spectroscopy of massive black holes with the space interferometer LISA},
  journal      = {Phys. Rev. D},
  volume       = {73},
  number       = {6},
  pages        = {064030},
  year         = {2006},
  doi          = {10.1103/PhysRevD.73.064030}
}

@article{Bhagwat2020,
  author       = {Bhagwat, S. and Okounkova, M. and Giesler, M. and Johnson-McDaniel, N. and others},
  title        = {Detectability of black hole quasinormal modes in current and future gravitational wave detectors},
  journal      = {Phys. Rev. D},
  volume       = {101},
  number       = {10},
  pages        = {104032},
  year         = {2020},
  doi          = {10.1103/PhysRevD.101.104032}
}

@misc{chandrasekhar1985mathematical,
  title={The mathematical theory of black holes},
  author={Chandrasekhar, Subrahmanyan and Thorne, Kip S},
  year={1985},
  publisher={American Association of Physics Teachers}
}

@article{yang2025black,
  title={Black hole quasinormal mode resonances},
  author={Yang, Yiqiu and Berti, Emanuele and Franchini, Nicola},
  journal={arXiv preprint arXiv:2504.06072},
  year={2025}
}

@inproceedings{jacot2018neural,
  author    = {Jacot, Arthur and Gabriel, Franck and Hongler, Cl\'ement},
  title     = {Neural Tangent Kernel: Convergence and Generalization 
               in Neural Networks},
  booktitle = {Advances in Neural Information Processing Systems (NeurIPS)},
  volume    = {31},
  year      = {2018},
  eprint    = {1806.07572},
}

@inproceedings{finn2017model,
  author    = {Finn, Chelsea and Abbeel, Pieter and Levine, Sergey},
  title     = {Model-Agnostic Meta-Learning for Fast Adaptation 
               of Deep Networks},
  booktitle = {Proceedings of the 34th International Conference 
               on Machine Learning (ICML)},
  pages     = {1126--1135},
  year      = {2017},
  eprint    = {1703.03400},
}

@article{GWTC5,
  author        = "{LIGO Scientific Collaboration, Virgo Collaboration and KAGRA Collaboration}",
  title         = "{GWTC-5.0: Observations from the Second Part of the Fourth LIGO-Virgo-KAGRA Observing Run and Updates to the Gravitational-Wave Transient Catalog}",
  eprint        = "2605.27225",
  archivePrefix = "arXiv",
  primaryClass  = "gr-qc",
  doi           = "10.48550/arXiv.2605.27225",
  year          = "2026"
}

@article{saha2025revisiting,
  author        = {Saha, Sagnik and Silva, Hector O.},
  title         = {Quasinormal modes of {Kerr--Newman} black holes: Revisiting the {Dudley--Finley} approximation},
  journal       = {Phys. Rev. D},
  volume        = {113},
  pages         = {064009},
  year          = {2026},
  doi           = {10.1103/2wc1-yntl},
  eprint        = {2510.05354},
  archivePrefix = {arXiv},
  primaryClass  = {gr-qc}
}

@article{dudley1977separation,
  author  = {Dudley, Alan L. and Finley, III, J. D.},
  title   = {Separation of Wave Equations for Perturbations of General Type-{D} Space-Times},
  journal = {Phys. Rev. Lett.},
  volume  = {38},
  pages   = {1505--1508},
  year    = {1977},
  doi     = {10.1103/PhysRevLett.38.1505}
}

@article{dudley1979covariant,
  author  = {Dudley, Alan L. and Finley, III, J. D.},
  title   = {Covariant Perturbed Wave Equations in Arbitrary Type-{D} Backgrounds},
  journal = {J. Math. Phys.},
  volume  = {20},
  pages   = {311--328},
  year    = {1979},
  doi     = {10.1063/1.524064}
}

@misc{Zenodo,
  author       = {Pombo, Alexandre M.},
  title        = {{Kerr--Newman quasinormal-mode dataset: coupled
                   gravito-electromagnetic modes from a hybrid
                   SpectralPINN solver}},
  year         = {2026},
  publisher    = {Zenodo},
  doi          = {10.5281/zenodo.21359120},
  note         = {\href{https://doi.org/10.5281/zenodo.21359120}{doi:10.5281/zenodo.21359120}}
}

@article{GWTC4,
  author        = "{LIGO Scientific Collaboration, Virgo Collaboration and KAGRA Collaboration}",
  title         = "{GWTC-4.0: Updating the Gravitational-Wave Transient Catalog with Observations from the First Part of the Fourth LIGO-Virgo-KAGRA Observing Run}",
  eprint        = "2508.18082",
  archivePrefix = "arXiv",
  primaryClass  = "gr-qc",
  year          = "2025"
}

@article{kokkotas1988black,
  title={Black-hole normal modes: A WKB approach. III. The Reissner-Nordstr{\"o}m black hole},
  author={Kokkotas, Kostas D and Schutz, Bernard F},
  journal={Physical Review D},
  volume={37},
  number={12},
  pages={3378},
  year={1988},
  publisher={APS}
}

@article{leaver1990quasinormal,
  title={Quasinormal modes of Reissner-Nordstr{\"o}m black holes},
  author={Leaver, Edward W},
  journal={Physical Review D},
  volume={41},
  number={10},
  pages={2986},
  year={1990},
  publisher={APS}
}

@article{leaver1985analytic,
  title={An analytic representation for the quasi-normal modes of Kerr black holes},
  author={Leaver, Edward W},
  journal={Proceedings of the Royal Society of London. A. Mathematical and Physical Sciences},
  volume={402},
  number={1823},
  pages={285--298},
  year={1985},
  publisher={The Royal Society London}
}

@article{teukolsky1973perturbations,
  title={Perturbations of a rotating black hole. I. Fundamental equations for gravitational, electromagnetic, and neutrino-field perturbations},
  author={Teukolsky, Saul A},
  journal={Astrophysical Journal, Vol. 185, pp. 635-648 (1973)},
  volume={185},
  pages={635--648},
  year={1973}
}

@article{luna2023solving,
  title={Solving the Teukolsky equation with physics-informed neural networks},
  author={Luna, Raimon and Calder{\'o}n Bustillo, Juan and Seoane Mart{\'\i}nez, Juan Jos{\'e} and Torres-Forne, Alejandro and Font, Jose A},
  journal={Physical Review D},
  volume={107},
  number={6},
  pages={064025},
  year={2023},
  publisher={APS}
}

@book{MasonHandscomb2003,
  author    = {Mason, J. C. and Handscomb, D. C.},
  title     = {Chebyshev Polynomials},
  publisher = {Chapman \& Hall/CRC},
  year      = {2003}
}

%
\appendix
%
%
    \section{Dataset Heatmap} \label{SA}
%
    This appendix collects the complete set of KN QNM frequency surfaces computed in this work. For each mode we show −Im($\omega M$) (top row) and Re($\omega M$) (bottom row) as interpolated heatmaps over the sub-extremal parameter space, with contour lines at fixed frequency intervals.

    \begin{figure}[h!]
	   \centering
	   \includegraphics[width=1\textwidth]{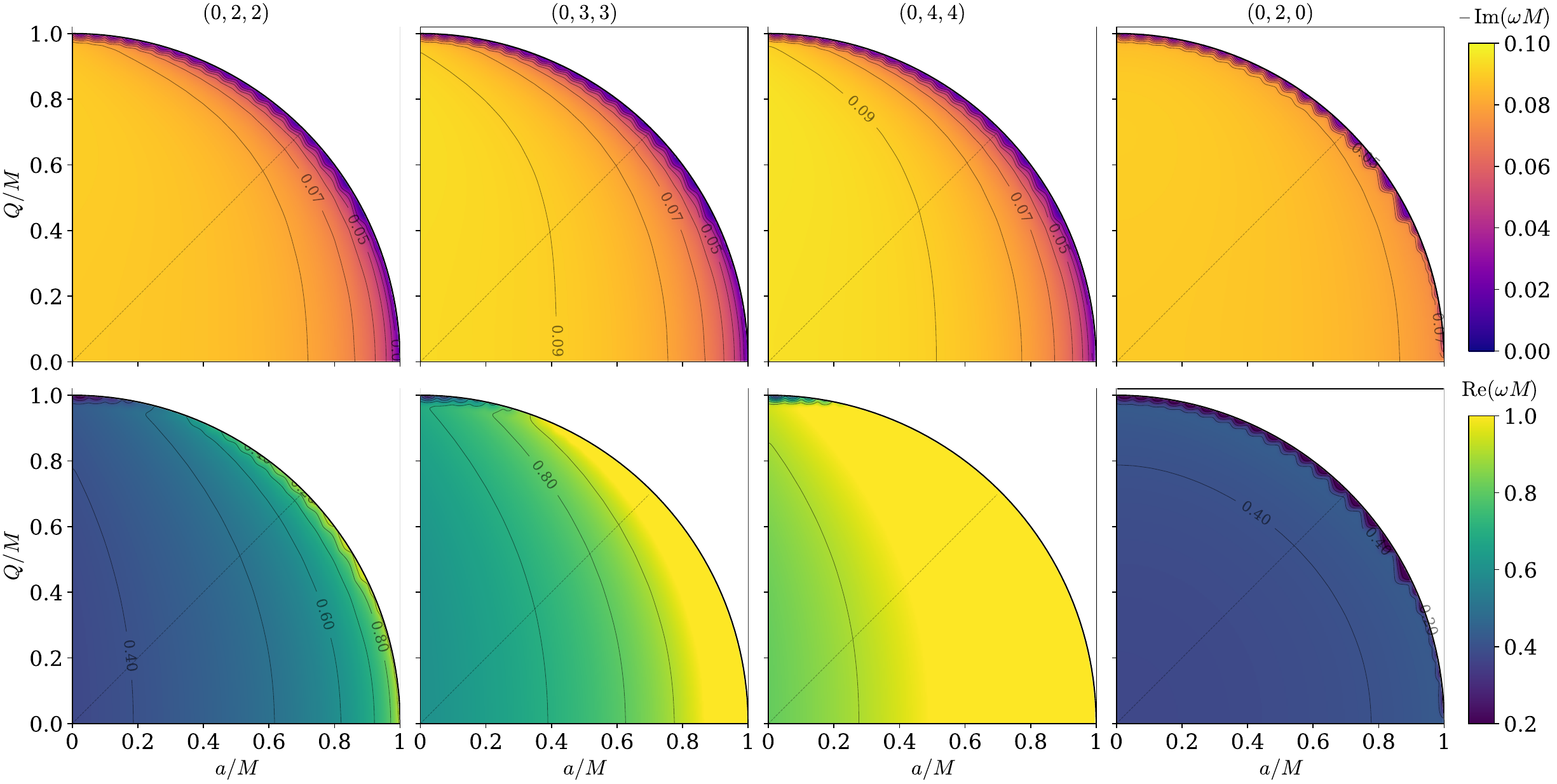}\\
      \caption{Photon-sphere Grav-led QNM frequency surfaces over the sub-extremal $(a/M,\,Q/M)$ parameter space for, from left to right, the $(n,\ell,m)=(0,2,2)$, $(0,3,3)$, $(0,4,4)$ and $(0,2,0)$ modes. (Top) damping rate $-\mathrm{Im}(\omega M)$; (Bottom) oscillation frequency $\mathrm{Re}(\omega M)$. Contour lines are drawn at fixed frequency intervals. The domain is bounded from above by the extremality curve $a^2/M^2+Q^2/M^2=1$.}
	   \label{F8}
    \end{figure}

    \begin{figure}[h!]
	   \centering
	   \includegraphics[width=1\textwidth]{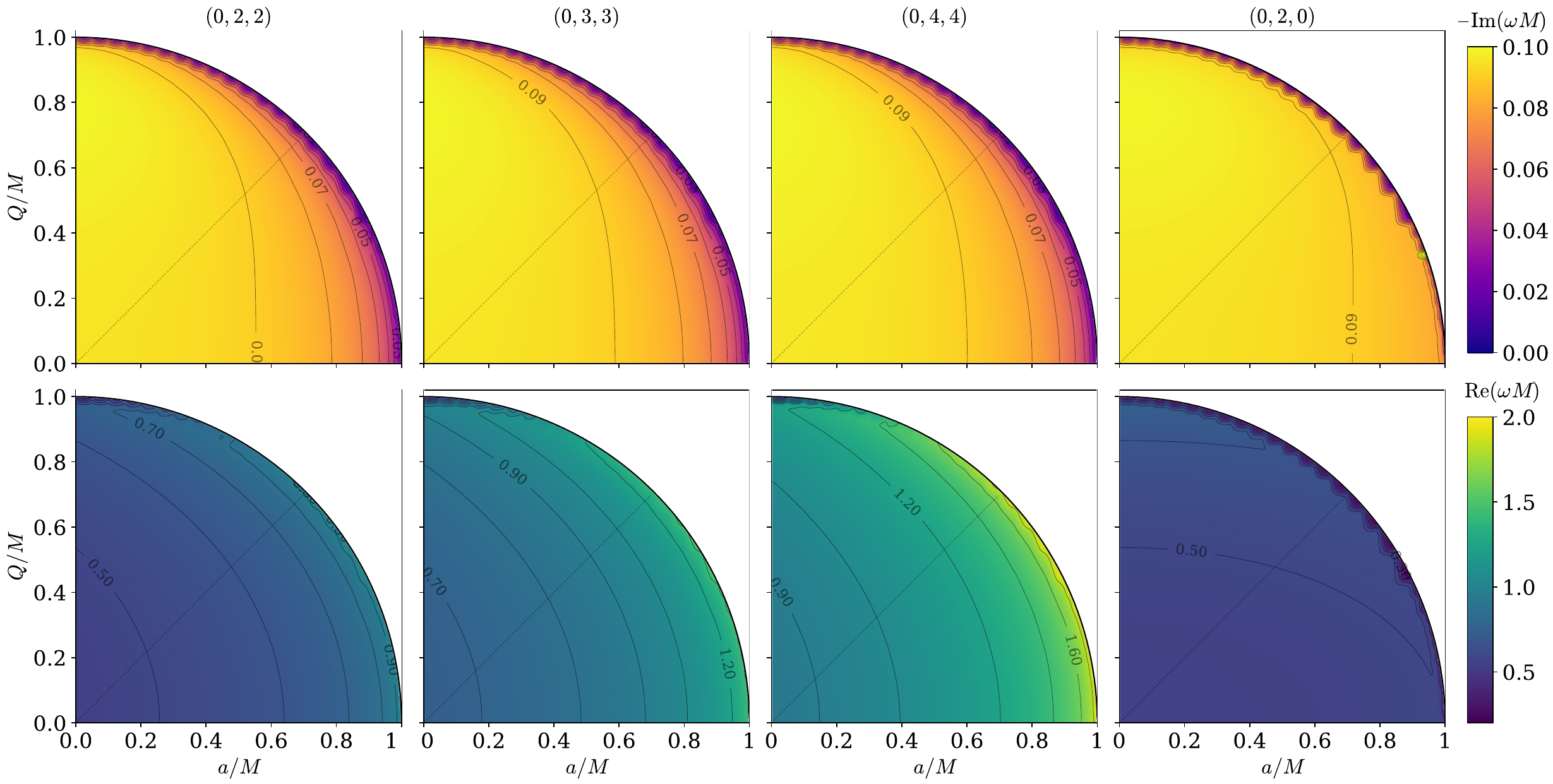}\\
      \caption{Photon-sphere Vect-led QNM frequency surfaces over the sub-extremal $(a/M,\,Q/M)$ parameter space for, from left to right, the $(n,\ell,m)=(0,2,2)$, $(0,3,3)$, $(0,4,4)$ and $(0,2,0)$ modes. (Top) damping rate $-\mathrm{Im}(\omega M)$; (Bottom) oscillation frequency $\mathrm{Re}(\omega M)$. Contour lines are drawn at fixed frequency intervals. The domain is bounded from above by the extremality curve $a^2/M^2+Q^2/M^2=1$.}
	   \label{F9}
    \end{figure}

    \begin{figure}[h!]
	   \centering
	   \includegraphics[width=1\textwidth]{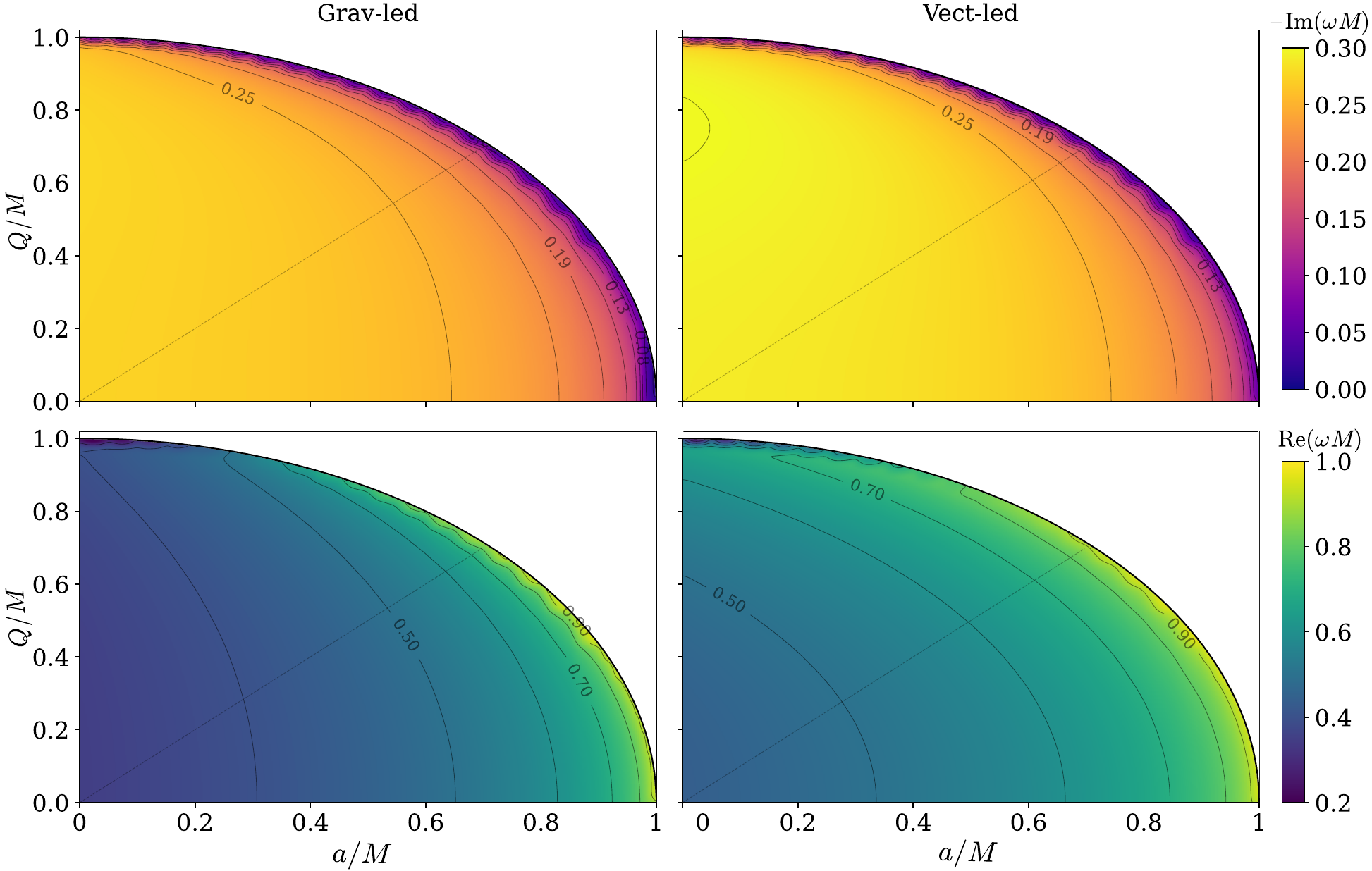}\\
      \caption{Photon-sphere $(n,\ell,m)=(1,2,2)$ overtone QNM frequency surfaces over the sub-extremal $(a/M,\,Q/M)$ parameter space for the (Left) Grav-led branch; (Right) Vect-led branch. (Top) damping rate $-\mathrm{Im}(\omega M)$; (Bottom) oscillation frequency $\mathrm{Re}(\omega M)$. Contour lines are drawn at fixed frequency intervals. The domain is bounded from above by the extremality curve $a^2/M^2+Q^2/M^2=1$.}
	   \label{F10}
    \end{figure}

    \begin{figure}[h!]
	   \centering
	   \includegraphics[width=1\textwidth]{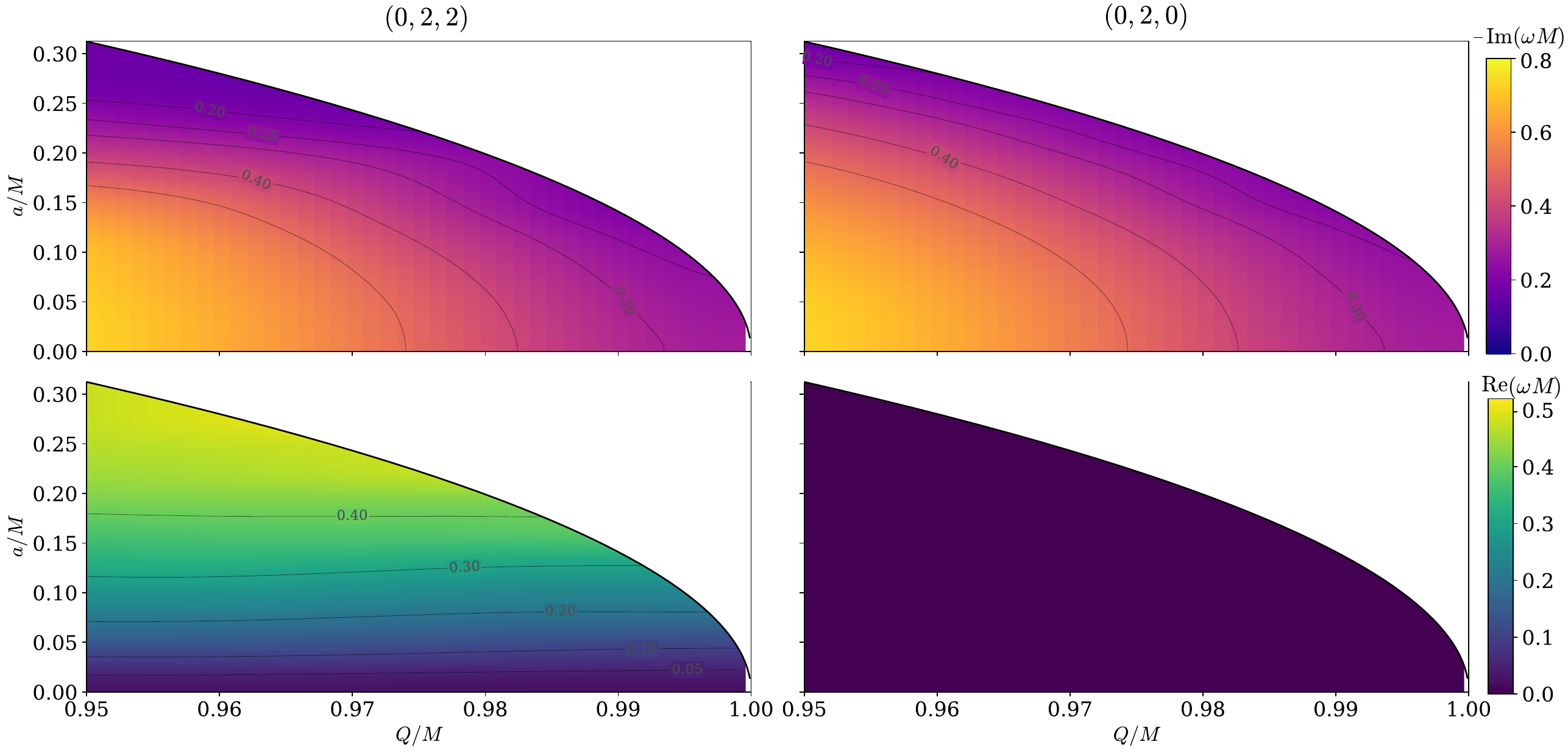}\\
      \caption{Near-horizon Grav-led QNM frequency surfaces for the $(n,\ell,m)=(0,2,2)$ (left) and $(0,2,0)$ (right) modes. (Top) damping rate $-\mathrm{Im}(\omega M)$; (Bottom) oscillation frequency $\mathrm{Re}(\omega M)$. Contour lines are drawn at fixed frequency intervals.}
	   \label{F11}
    \end{figure}

\end{document}